\documentstyle[11pt,amssymb,times,epsf]{article}
\newcommand{\be}{\begin{equation}}
\newcommand{\ee}{\end{equation}}
\newcommand{\ba}{\begin{array}}
\newcommand{\ea}{\end{array}}
\newcommand{\bea}{\begin{eqnarray}}
\newcommand{\eea}{\end{eqnarray}}

\renewcommand{\l}{\newline\null}

\newcommand{\lrar}{\longrightarrow}
\newcommand{\rar}{\rightarrow}
\newcommand{\p}{\partial}
\newcommand{\ol}{\overline}
\newcommand{\ti}{\tilde}
\newcommand{\la}{\langle}
\newcommand{\ra}{\rangle}
\def\figskip{\vskip .5cm plus 3mm minus 2mm}
\def\hbar{h\!\!\!/}

\begin{document}
\begin{titlepage}
June 1994 (revised March 1995)\hfill PAR-LPTHE 94/24
\begin{flushright} hep-ph/9406349 \end{flushright}
\vskip 3.5cm
\begin{center}
{\bf COMMENTS ON THE STANDARD MODEL OF ELECTROWEAK INTERACTIONS.}
\end{center}
\smallskip
\centerline{B. Machet,
     \footnote[1]{Member of `Centre National de la Recherche Scientifique'.}
     \footnote[2]{E-mail: machet@lpthe.jussieu.fr}
     }
\vskip 5mm
\centerline{{\em Laboratoire de Physique Th\'eorique et Hautes Energies},
     \footnote[3]{LPTHE, tour 16/1$^{er}$ \'etage,
          Universit\'e P. et M. Curie, BP 126, 4 place Jussieu,
          F 75252 PARIS CEDEX 05 (France).}
     }
\centerline{\em Universit\'es Pierre et Marie Curie (Paris 6) et Denis
Diderot (Paris 7);}\centerline{\em  Unit\'e associ\'ee au CNRS D0 280.}
\vskip 2cm
{\bf Abstract:} The Standard Model of electroweak interactions is shown to
include a gauge theory for the observed scalar and pseudoscalar mesons. This
is done by exploiting the consequences of embedding the $SU(2)_L\times U(1)$
group  into the chiral group of strong
interactions and by considering explicitly the Higgs boson and its three
companions inside the standard scalar 4-plet as composite.  No extra scale of
interaction is needed.
Quantizing by the Feynman path integral reveals how, in the
`Nambu Jona-Lasinio approximation', the quarks and the Higgs boson become
unobservable, and the theory anomaly-free. Nevertheless, the `anomalous'
couplings of the pseudoscalar mesons to gauge fields spring again from the
constraints associated with their compositeness itself.  This work is the
complement  of ref.~\cite{BellonMachet}, where  the leptonic sector was shown
to be compatible with a purely vectorial theory and, consequently, to be also
anomaly-free. The bond between quarks and leptons loosens.

\smallskip
{\bf PACS:} 11.15.Ex, 11.30.-j, 11.30.Rd, 11.40.Ha, 12.15.-y, 12.15.Ff,
12.38.Cy, 12.38.Lg, 12.60.Nz, 12.60.Rc, 13.20.-v, 13.40.Hq, 14.40.-n,
14.80.Bn, 14.80.Cp.
\vfill
\null\hfil\epsffile{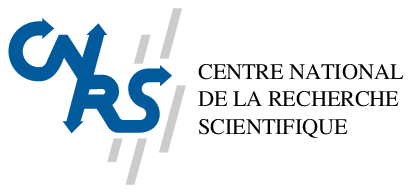}
\end{titlepage}

\section{Introduction.}

Leaving aside gravitation, the interactions of fermions split into two
categories: strong and electroweak; no true unification is yet achieved
between the two.
In the work \cite{BellonMachet}, we showed that  the effective $V-A$ structure
of leptonic weak interactions can be deduced from a purely vectorial theory,
and that the `right-handed' neutrino becomes unobservable as  asymptotic state.
I focus now
on the quark sector, where both types of interactions are at work.
I present an attempt to partially fill the gap between them, with the
(restricted) meaning that some characteristics, attributed before either to
one or to the other, are now visualized from a unified point of view.

Quantum Chromodynamics, the would-be candidate for a theory of strong
interactions of quarks and gluons  (see for example \cite{MarcianoPagels}),
 will not be dealt with here. It fails, up
to now, to describe the basics of low energy interactions between
hadrons, which cannot themselves be accounted for as asymptotic states. As
the point cannot be settled whether this is a failure of the theory or a
result of our incapacity to compute, I will consider QCD to be irrelevant
at low energy.

Going in some way backwards in time, I will instead be concerned here with
the link between one fundamental aspect underlying strong interactions,
chiral symmetry (see for example \cite{AdlerDashen}),
 including the diagonal `flavour group' at the root of the
constituent quark model, and the gauge group of electroweak
interactions \cite{GlashowSalamWeinberg}. I
give a new description of the breaking of chiral symmetry, leading in
particular to a new interpretation of the associated Goldstone bosons
\cite{Goldstone}; the
gauge group being identified with a subgroup of the chiral group of strong
interactions, the latter become now electroweak eigenstates, mixing, into the
same multiplets,
scalar and pseudoscalar mesons: electroweak and `strong' eigenstates are
linked together by relations which generalize those existing, for example,
between the $K_1, K_2$ mesons and $K^0, \bar K^0$.
This identification has numerous
consequences, only a part of which is sketched here; I will
focus on the resulting classification of the scalar and
pseudoscalar mesons into $SU(2)_L\times U(1)$ multiplets, which yield new
mass relations and degeneracies. In particular the question of the mass of
the $\eta'$ meson can find a natural solution outside the intricacies of the
QCD vacuum \cite{Coleman77}.

By naming the $\eta'$ without adding any `techni-'like  prefix, I already mean
that the above mentioned scalar and pseudoscalar fields are directly related
to ---in fact just linear combinations of--- the daily observed scalar and
pseudoscalar mesons, and not to a higher scale of interactions like in
technicolour theories \cite{SusskindWeinberg}. The accent is specially put
on two processes: the leptonic decays of the pseudoscalar mesons and their
`anomalous' couplings to two gauge fields.

Unlike in the traditional scheme of chiral symmetry breaking, we do not
expect $N^2$ ($N =\ $ number of flavours of quarks) `light' particles.
 In general, mesons can be given masses in an $SU(2)_L\times U(1)$ invariant
way, without any reference to the parameters called `quark masses', and only
three among the above electroweak eigenstates become the longitudinal
components of the three massive gauge fields. The different mass scales
appear directly at the mesonic level in the Lagrangian.

Another original aspect of this model is the relation between fermions and
mesons that results from its quantization. Nature shows, in an, up to now,
unambiguous way, that they do not coexist as asymptotic states; this
is akin to saying
that quarks, as asymptotic states, have infinite masses.
This is what occurs here, by the Feynman path integral quantization,
when, in the so-called `Nambu Jona-Lasinio' approximation
\cite{NambuJonaLasinio}, corresponding to keeping the leading order in a
development in inverse powers of the number of flavours,
 the three longitudinal components of the massive $W$'s  and the Higgs
boson (forming the usual scalar sector) are explicitly considered to be
composite. Two ingredients concur for that:\l
- first, the necessity, when integrating on both
fermions and bound states which are non-independent degrees of freedom,
to explicitly introduce constraints in the path integral; they can be
exponentiated into an effective Lagrangian;\l
- secondly,
the freezing of the fermionic degrees of freedom themselves when the Higgs
boson gets a non-vanishing vacuum expectation value: an infinite mass for
the quarks appears in the above effective Lagrangian. In addition, the Higgs
boson becomes itself infinitely massive and unobservable.

The result is a gauge, unitary theory of mesons and gauge fields, the
latter being as usual the massless photon and the three massive $W$'s.
I show that it is anomaly-free but that, nevertheless, `anomalous' couplings
of the pseudoscalar mesons to gauge fields are
rebuilt from the constraints, with subtleties which will be evoked.

The mass of the Higgs boson being infinite, the massive gauge bosons are
expected to be strongly interacting \cite{Veltman}, which is just another
facet of strong interactions that usual pseudoscalar and vector mesons
undergo. A bridge thus starts being built between the weak and strong sector.

Due to the complexity of the effective Lagrangian originating from the
constraints, I only study renormalizability in the already mentioned
`Nambu Jona Lasinio approximation' \cite{NambuJonaLasinio},
in which  precisely the sole bound states propagate.

Similarly, for the sake of simplicity, I only study the 4-flavour
case, leaving temporarily aside phenomena like the violation of $CP$, and the
largest part of the argumentation is performed in an abelian model which has
all the characteristic and interesting features without the unnecessary
intricacies linked with a non-abelian group.

The title of this work is justified by the conservative point of
view adopted here. The only {\em addenda} with respect to the
Glashow-Salam-Weinberg model are mainly conceptual, since they concern the
embedding of the gauge group into the chiral group, and  considering the
standard scalar multiplet, {\em i.e.} the Higgs boson and its three companions,
as composite. The consequences are, despite what one could think, large, and
not so standard, and have effects on many basic aspects of the electroweak
interactions of mesons. By establishing a  link between two, up to now
disconnected, groups of symmetry, one hopes that they will provide a new
insight into a possible unification with the strong interactions.

\section{Embedding the gauge group into the chiral group.}

\subsection{The `standard' picture of chiral symmetry breaking.}

Let $N$ be the number of quark flavours. The group of chiral symmetry is
$ {\cal G}_\chi = U(N)_L \times U(N)_R$, where the subscripts `L' and `R'
mean `left' and `right' respectively.
In the standard picture \cite{AdlerDashen,deAlfaroFubiniFurlan},
 it is broken down to the diagonal $U(N)$ of
flavour, which, in the $N=3$ case, includes the $SU(3)$ of Gell-Mann
\cite{GellMann}.  The $N^2$
corresponding Goldstone bosons are identified with the $N^2$ pseudoscalar
mesons. In the quark model, chiral symmetry breaking is triggered by `quark
condensation'
\be
\la \ol\Psi\Psi\ra = N\mu^3.
\ee
The non-zero and very different masses of the `Goldstones' are accounted
for by putting by hand different quark masses in the Lagrangian,
which produce an explicit breaking of chiral symmetry.
Low energy theorems combined with the
`Partially Conserved Axial Current' (PCAC) hypothesis, link quark masses $m$,
the scale of breaking $\mu$, the mass of the Goldstones $M_G$ and their
weak decay constants $f_G$ by relations of the type
\cite{GellMannOakesRennerDashen}
\be
m\: \mu^3 = -\;\kappa\;f_G^2 M_G^2,
\label{eq:DASHEN}\ee
where $\kappa$ is a numerical group factor of the order of unity.
The fact that the $\eta'$ meson has a much heavier mass than the pion has
been a long standing problem \cite{Coleman77}.

\subsection{The standard picture of electroweak symmetry breaking.}

The Standard Model of electroweak interactions by Glashow, Salam and Weinberg,
\cite{GlashowSalamWeinberg}\cite{AbersLee}
involves four (real) scalar fields, including the Higgs boson. The gauge
symmetry is broken by the latter acquiring a non-vanishing vacuum expectation
value in the vacuum
\be
\la H\ra = v/\sqrt{2}.
\ee
The three companions of the Higgs are the Goldstone bosons of the broken
$SU(2)$ symmetry; they can be gauged into the third polarizations
of the three massive gauge fields, the $W$ gauge bosons. The zero mass of the
photon is preserved, corresponding to an unbroken $U(1)$ group for pure
electromagnetism.

The coupling of the gauge fields to the quarks of electric charge $-1/3$
occurs through a mixing matrix, the Cabibbo matrix \cite{Cabibbo}
 in the 4-quark case
(generalized to the Kobayashi-Maskawa matrix \cite{KobayashiMaskawa}
 in the 6-quark case).
The introduction of the Cabibbo mixing angle being one of the most well
confirmed fact, I will not question its existence nor that of a unitary
mixing matrix.

\subsection{The embedding.}

The gauge group ${\cal G}_s = SU(2)_L\times U(1)$ acts on the fermions
(quarks) and on the gauge fields.
I recall that, for the sake of simplicity, we only deal here with the 4-quark
case. The reader can understand how the generalization proceeds.

\subsubsection{Identifying the generators of the gauge group as
$\bf 4\times 4$ matrices of $\bf U(4)$.}

We work in the fermion basis
\be
\Psi = \left(\begin{array}{c} u  \\
                              c  \\
                              d  \\
                              s  \\ \end{array} \right).
\ee
The identification of the generators of ${\cal G}_s$ as $4\times 4$ matrices
  proceeds as follows:\l
\be
{\Bbb T}_L^+={\Bbb T}_L^1+i{\Bbb T}_L^2={{1-\gamma_5}\over 2}\ {\Bbb C},\
{\Bbb T}_L^-={\Bbb T}_L^1-i{\Bbb T}_L^2={{1-\gamma_5}\over 2}\ {\Bbb C}^{\dag},
\ {\Bbb T}_L^3={1\over2}{{1-\gamma_5}\over2}\ {\Bbb N};
\ee
\quad - for $U(1)_L$:
\be
{\Bbb Y}_L={{1-\gamma_5}\over 2}\ {\Bbb Y}={1\over 6}
{{1-\gamma_5}\over 2}\ {\Bbb I};
\ee
\quad - for $U(1)_R$:
\be
{\Bbb Y}_R={{1+\gamma_5}\over 2}\ {\Bbb Y}={{1+\gamma_5}\over 2}\ {\Bbb Q},
\ee
where
\be
{\Bbb C}=\left(\begin{array}{ccc}
                        0 & \vline & {\bf C}\\
                        \hline
                        0 & \vline & 0           \end{array}\right);
{\Bbb N}=\left(\begin{array}{ccc}
                        1 & \vline & 0\\
                        \hline
                        0 & \vline & {-1}        \end{array}\right);
{\Bbb Q}=\left(\begin{array}{ccc}
                        {2\over 3} & \vline & 0\\
                        \hline
                        0 & \vline & -{1\over 3}  \end{array}\right).
\ee
${\bf C}$ is the customary $2\times 2$ Cabibbo mixing matrix \cite{Cabibbo}.
$\Bbb I$ is the unit matrix.
We have the usual Gell-Mann-Nishijima relation
\be
{\Bbb Y}={\Bbb Y}_R +{\Bbb Y}_L= {\Bbb Q}-{\Bbb T}^3_L={\Bbb Q}_R + {\Bbb Y}_L.
\label{eq:GMN}
\ee
It is an essential feature that not only the ${\Bbb T}$'s and the right and
left $\Bbb Y$'s have the usual $SU(2)_L\times U(1)$ commutation
relations, but also that the ${\Bbb T}_L$'s and ${\Bbb Y}_L$  form a {\em
matrix  algebra} by themselves: any product of two among the four stays
in the set (this property makes the `left' part of the gauge group
isomorphic to $U(2)$). We have indeed:
\be\ba{cccccccccccc}
\{3{\Bbb Y}_L,{\Bbb T}^3_L\} &=& {\Bbb T}^3_L &,&
\{3{\Bbb Y}_L,{\Bbb T}^+_L\} &=& {\Bbb T}^+_L &,&
\{3{\Bbb Y}_L,{\Bbb T}^-_L\} &=& {\Bbb T}^-_L &,  \\
\{{\Bbb T}^3_L,{\Bbb T}^3_L\}&=& 3{\Bbb Y}_L  &,&
\{{\Bbb T}^3_L,{\Bbb T}^+_L\}&=& 0            &,&
\{{\Bbb T}^3_L,{\Bbb T}^-_L\}&=& 0            &,  \\
& & & &
\{{\Bbb T}^+_L,{\Bbb T}^-_L\} &=& 6{\Bbb Y}_L &.&
& & &
\ea
\ee
This property allows scalar representations of composite
quark-antiquark fields, mixing scalars and pseudoscalars; indeed, because of
the $\gamma_5$ matrix appearing in the `left' gauge generators, their action on
composite fermion operators of the form $\ol \Psi {\cal O} \Psi$ will involve
both their commutators and their anticommutators with $\cal O$.
It will be thoroughly
exploited in the rest of the paper; for the moment, we can
immediately see that the composite scalar multiplet
\be
\Phi = (H,\phi^3,\phi^+,\phi^-)= {v\over N\mu^3}
\ol\Psi
\left(
{1\over\sqrt{2}}{\Bbb I},
{1\over\sqrt{2}}\gamma_5{\Bbb N}, \gamma_5{\Bbb C},\gamma_5{\Bbb C}^{\dag}
\right)
\Psi,
\label{eq:PHI}\ee
where the conventions are
\be
\phi^+={\phi_1+i\phi_2\over\sqrt{2}},\
\phi^-={\phi_1-i\phi_2\over\sqrt{2}},
\ee
is isomorphic to the standard scalar multiplet of the Glashow-Salam-Weinberg
model. Here, $H$ is real and the $\phi$'s purely imaginary; we shall also
use the real fields $\vec\varphi$ such that $\vec\phi = i\;\vec\varphi$.
The action of ${\cal G}_s$ on $\Phi$, deduced from that on the fermions,
reads:
\bea
{\Bbb T}^i_L. \phi_j & = &
-{1\over 2}(i\;\varepsilon_{ijk}\phi_k + \delta_{ij}H) \\
{\Bbb T}^i_L. H & = & -{1\over 2} \phi_i,
\label{eq:ACTIONPHI}\eea
showing that it is as usual the sum of two representations $1/2$ of
$SU(2)_L$. The action of $U(1)$ is deduced from the
Gell-Mann-Nishijima relation and that, trivial of the charge operator $\Bbb
Q$. This means in particular that the mechanism of mass generation for the
gauge bosons stays unaltered.

By this embedding, chiral and gauge symmetry breaking appear naturally as
two aspects of the same phenomenon, since
\be
\la H \ra = v/\sqrt{2}
\ee
is now equivalent to
\be
\la \ol\Psi \Psi \ra = N\mu^3.
\ee
It is {\em not} our goal here to trace the origin of this phenomenon, that
we take for granted and consider to be a constraint on the system. We shall
only
study its consequences and show its consistency with  the existence of bound
states. This makes our work much less ambitious, in particular, than that of
Nambu and Jona-Lasinio \cite{NambuJonaLasinio}, the analogies and
differences with which we shall stress in the following (see
section~\ref{sec:NJL}).

Strictly speaking, we could at that point proceed with the study of the
Standard Model, taking only into account the modification due to the
compositeness of the scalar multiplet. We shall indeed show later that
the pseudoscalar partners of the Higgs boson behave like linear combinations
of some of the {\em observed} pseudoscalar mesons, and {\em not} like
`technimesons' which would correspond to another higher scale of interactions
\cite{SusskindWeinberg} (and I will show that the latter is not needed here).
However, $2N^2 -3$ scalars and
pseudoscalars composed from a quark-antiquark pair are still missing; this is
why I will first continue to exploit the very peculiar group structure of
the model and show how the other mesons fit into this framework.
I stay at the level of simple group theory arguments, postponing dynamical
considerations to a later part of the paper.

\subsection{More structure.}

I show that the existence of another scalar 4-plet of the group ${\cal G}_s$
is related with the presence of a $\left(SU(2)_L\times U(1))\times
(SU(2)_L\times U(1)\right)$ group of symmetry.
The mixing angle appears as controlling
the embedding of ${\cal G}_s$ into the latter.

Let ${\cal P}_1$ and ${\cal P}_2$ be the $2\times 2$ orthogonal projector
matrices
\be
{\cal P}_1 = {1\over 2}\left( \ba{rr} 1 & i \cr
                     -i & 1 \ea \right), \qquad
{\cal P}_2 = {1\over 2}\left( \ba{rr} 1 & -i\cr
                      i & 1 \ea \right);
\ee
they satisfy
\be
{\cal P}_1^2 = {\cal P}_1,\quad{\cal P}_2^2 = {\cal P}_2,\quad
{\cal P}_1 {\cal P}_2 = {\cal P}_2 {\cal P}_1 = 0.
\ee
The 6 matrices built from the ${\cal P}$'s
\be
t^+ = \left( \ba{ccc} 0 &\vline & {\cal P} \cr
                    \hline 0 &\vline & 0 \ea \right),\quad
t^- = \left( \ba{ccc} 0 & \vline & 0 \cr
                \hline {\cal P} & \vline & 0 \ea \right),\quad
t^3 = {1\over 2} \left( \ba{ccc} {\cal P} &\vline & 0 \cr
                                     \hline 0 &\vline & -{\cal P} \ea\right)
\ee
form the generators of two $SU(2)$ commuting groups, the first associated
with ${\cal P}_1$, the second associated with ${\cal P}_2$. We introduce
in addition the 4 matrices $n_1,n_2, q_1, q_2$
according to
\be
n = \left( \ba {ccc} \cal P &\vline & 0 \cr
                            \hline  0 &\vline & \cal P \ea\right),\quad
q = \left( \ba {ccc} {2\over 3} \cal P &\vline & 0 \cr
                          \hline  0 &\vline & -{1\over 3}\cal P \ea\right),
\ee
to which we associate the `hypercharge' generators
\be
y_L = {1\over 6} n_L, \quad y_R = q_R.
\ee
$({\vec t_{L1}}, y_1)$ form a first $SU(2)_L \times U(1)$ group,
${\cal G}_1$, $({\vec t_{L2}}, y_2)$ form a second \hbox{$SU(2)_L \times
U(1)$ group, ${\cal G}_2$.}

The commutation relations of any $SU(2)$ group are unaltered if we perform a
rotation between the $t^1$ and $t^2$ generators, or, equivalently, if we
replace $t^+$ by $e^{i\alpha}\;t^+$ and $t^-$ by $e^{-i\alpha}\; t^-$,
leaving $t^3$ unchanged; we call the corresponding group ${\cal G}(\alpha)$.
It is now easy to see that the group of the Standard Model, ${\cal G}_s$ can
be written symbolically
\be
{\cal G}_s = {\cal G}_1(-\theta_c) + {\cal G}_2(\theta_c),
\ee
where $\theta_c$ is the Cabibbo angle. By the above formula, I mean  the
following relations between the generators of the groups:
\bea
{\Bbb T}^3 &=& t_1^3 + t_2^3,
\ {\Bbb T}^+ = e^{-i\theta_c}t_1^+ + e^{i\theta_c} t_2^+,
\ {\Bbb T}^- = e^{i\theta_c}t_1^- + e^{-i\theta_c}t_2^-,\cr
{\Bbb Y}_L &=& y_{1L} + y_{2L},\ {\Bbb Y}_R = y_{1R} + y_{2R}.
\eea
 $\theta_c$ now controls the
embedding of the gauge group into the product of two $SU(2)_L \times
U(1)$'s.

\subsubsection{Another scalar multiplet.}

Let us call $\Phi_1$ and $\Phi_2$ the two scalar multiplets
\be
\Phi_1 = {\sigma_1\over N} \ol \Psi(n_1,
                      \gamma_5 {\vec t_1})\Psi,\quad
\Phi_2 = {\sigma_2\over N} \ol\Psi( n_2,
                      \gamma_5 {\vec t_2})\Psi.
\label{eq:PHI1PHI2}\ee
$\sigma_1$ and $\sigma_2$ have dimension $mass^{-2}$.
${\cal G}_1$ has a vanishing action on $\Phi_2$, and ${\cal G}_2$ on
$\Phi_1$.

It is easy to check that we obtain other representations  by performing the
same rotations on the generators appearing in the $\Phi$'s as in the above
subsection; more precisely, for example,
\be
\Phi_1(\alpha) = {\sigma\over N}
\ol\Psi\left(
 {1\over\sqrt{2}}n_1,
{1\over\sqrt{2}}\gamma_5 t_1^3,
e^{i\alpha}\gamma_5 t_1^+,
e^{-i\alpha}\gamma_5 t_1^-
\right)\Psi
\ee
is a representation of ${\cal G}_1(\alpha)$, invariant by ${\cal G}_2$.
We can now identify the basic representation $\Phi$ as
\be
\Phi = \Phi_1(-\theta_c) + \Phi_2(\theta_c),
\ee
and find another scalar multiplet representation of ${\cal G}_s$
\bea
\Xi &=& \Phi_1(-\theta_c) - \Phi_2(\theta_c) = \cr & &
\hskip -1cm{\xi\over N}\ol\Psi\left(
{1\over\sqrt{2}}(n_1 -n_2),
{1\over\sqrt{2}}\gamma_5 (t_1^3 -t_2^3),
 \gamma_5 (e^{-i\theta_c}t_1^+ -e^{i\theta_c} t_2^+),
 \gamma_5 (e^{i\theta_c}t_1^- -e^{-i\theta_c} t_2^-)
\right)\Psi.
\label{eq:XI}\eea
$\xi$ has dimension $mass^{-2}$.
Note that this occurs despite the fact that ${\cal G}_1(-\theta_c) -{\cal
G}_2(\theta_c)$ is {\em not} a group. The structure at the origin of
$\Xi$ is actually a $Z(2)$ gradation of the gauge group.

To have a more explicit view in terms of quarks, note that
\be
e^{-i\theta_c} {\cal P}_1 - e^{i\theta_c}{\cal P}_2 =
-i\left(\ba{cc} \sin\theta_c & -\cos\theta_c \cr
                \cos\theta_c &  \sin\theta_c  \ea\right);\quad
{\cal P}_1 - {\cal P}_2 =
\left(\ba{cc}  0 & i \cr
              -i & 0 \ea\right).
\ee

\subsection{The last multiplets.}

With four flavours of quarks, we expect sixteen scalar and sixteen pseudoscalar
fermion-antifermion pairs, of which we have up to now exhibited only eight.
I now show how the last twenty-four appear.

\subsubsection{Two more scalar 4-plets.}

Let ${\cal P}^+$ and ${\cal P}^-$ be the two $2\times 2$ matrices
\be
{\cal P}_+ = {1\over 2}\left( \ba{rr} i  & 1 \cr
                            1  & -i \ea \right), \quad
{\cal P}_- = {1\over 2}\left( \ba{rr} -i & 1 \cr
                             1 & i \ea \right).
\ee
They are nilpotent:
\be
{{\cal P}_+}^2 = {{\cal P}_-}^2 =0.
\ee
We have the relations
\bea
& &{\cal P}_+ {\cal P}_- = {\cal P}_1,\ {\cal P}_- {\cal P}_+ = {\cal P}_2,\cr
& &{\cal P}_1 {\cal P}_+ = {\cal P}_+ {\cal P}_2 = {\cal P}_+,\cr
& &{\cal P}_2 {\cal P}_- = {\cal P}_- {\cal P}_1 = {\cal P}_-,\cr
& &{\cal P}_1 {\cal P}_- = {\cal P}_- {\cal P}_2
       = {\cal P}_2 {\cal P}_+ ={\cal P}_+ {\cal P}_1 =0.
\eea
The four $2\times 2$ matrices ${\cal P}_1, {\cal P}_2, {\cal P}_+, {\cal P}_-$
span a $U(2)$ algebra.

{}From the last two, we can build two more, complex-conjugate, 4-dimensional
multiplets $\Delta$ and $\Theta$, representations of ${\cal G}_s$, according to
\begin{eqnarray}
&&\Delta = {\rho\over N}\times \cr
&&\hskip -2cm\ol\Psi\left(
{1\over\sqrt{2}}
    \left(\ba{ccc} e^{-i\theta_c}{\cal P}_+ &\vline & 0 \cr
         \hline 0 &\vline & e^{+i\theta_c}{\cal P}_+ \ea\right) ,
{1\over\sqrt{2}}
    \gamma_5\left(\ba{ccc}  e^{-i\theta_c}{\cal P}_+ &\vline & 0\cr
         \hline 0 &\vline & -e^{+i\theta_c}{\cal P}_+ \ea\right) ,
 \gamma_5\left(\ba{ccc} 0 &\vline & {\cal P}_+ \cr
            \hline 0 &\vline & 0 \ea\right) ,
 \gamma_5\left(\ba{ccc} 0 &\vline & 0 \cr
            \hline {\cal P}_+ &\vline & 0 \ea\right)
\right)\Psi,\cr
&&
\label{eq:DELTA}
\end{eqnarray}
\bea
&&\Theta ={\omega\over N}\times \cr
&&\hskip -2cm\ol\Psi\left(
{1\over\sqrt{2}}
       \left(\ba{ccc} e^{+i\theta_c}{\cal P}_- &\vline & 0 \cr
               \hline 0 &\vline & e^{-i\theta_c}{\cal P}_- \ea\right) ,
{1\over\sqrt{2}}
       \gamma_5\left(\ba{ccc} e^{+i\theta_c}{\cal P}_-  &\vline & 0 \cr
               \hline 0 &\vline & -e^{-i\theta_c}{\cal P}_- \ea\right) ,
 \gamma_5\left(\ba{ccc} 0 &\vline & {\cal P}_- \cr
       \hline 0 &\vline & 0 \ea\right) ,
 \gamma_5\left(\ba{ccc} 0 &\vline & 0 \cr
        \hline {\cal P}_- &\vline & 0 \ea\right)
\right)\Psi.\cr
&&
\label{eq:THETA}
\eea
$\rho$ and $\omega$ have dimension $mass^{-2}$.

\subsubsection{A trivial doubling.}

Is is trivial to uncover the last four 4-plets of scalars by performing on
$\Phi, \Xi, \Delta, \Theta$ a $\gamma_5$ transformation.
We get from $\Phi$ (eq.~(\ref{eq:PHI}))
\be
\tilde\Phi = {\varsigma\over N} \ol\Psi\left(
 {1\over\sqrt{2}}\gamma_5 {\Bbb I},
 {1\over\sqrt{2}}{\Bbb N},
 {\Bbb C},
 {\Bbb C}^{\dag}
\right)\Psi,
\label{eq:PHITILDE}\ee
from $\Xi$ (eq.~(\ref{eq:XI}))
\be
\tilde\Xi ={\tilde\xi\over N} \ol\Psi\left(
 {1\over\sqrt{2}}(n_1 - n_2)\gamma_5 ,
 {1\over\sqrt{2}}(t_1^3 -t_2^3),
 (e^{-i\theta_c}t_1^+ -e^{i\theta_c}t_2^+),
 (e^{i\theta_c}t_1^- -e^{-i\theta_c} t_2^-)
\right)\Psi,
\label{eq:XITILDE}\ee
from $\Delta$ (eq.~(\ref{eq:DELTA}))
\bea
&&\tilde\Delta = {\tilde\rho\over N}\times\cr
&&\hskip -2cm\ol\Psi\left(
{1\over\sqrt{2}}
    \gamma_5\left(\ba{ccc} e^{-i\theta_c}{\cal P}_+ &\vline & 0 \cr
         \hline 0 &\vline & e^{+i\theta_c}{\cal P}_+ \ea\right) ,
{1\over\sqrt{2}}
    \left(\ba{ccc} e^{-i\theta_c}{\cal P}_+ &\vline & 0 \cr
         \hline 0 &\vline & -e^{+i\theta_c}{\cal P}_+ \ea\right) ,
 \left(\ba{ccc} 0 &\vline & {\cal P}_+ \cr
            \hline 0 &\vline & 0 \ea\right) ,
 \left(\ba{ccc} 0 &\vline & 0 \cr
            \hline {\cal P}_+ &\vline & 0 \ea\right)
\right)\Psi,\cr
&&
\label{eq:DELTATILDE}
\eea
and from $\Theta$ (eq.~(\ref{eq:THETA}))
\bea
&&\tilde\Theta ={\tilde\omega\over N}\times\cr
&&\hskip -2cm\ol\Psi\left(
{1\over\sqrt{2}}
       \gamma_5\left(\ba{ccc} e^{+i\theta_c}{\cal P}_- &\vline & 0 \cr
               \hline 0 &\vline & e^{-i\theta_c}{\cal P}_- \ea\right) ,
{1\over\sqrt{2}}
       \left(\ba{ccc} e^{+i\theta_c}{\cal P}_- &\vline & 0 \cr
               \hline 0 &\vline & -e^{-i\theta_c}{\cal P}_- \ea\right) ,
 \left(\ba{ccc} 0 &\vline & {\cal P}_- \cr
       \hline 0 &\vline & 0 \ea\right) ,
 \left(\ba{ccc} 0 &\vline & 0 \cr
        \hline {\cal P}_- &\vline & 0 \ea\right)
\right)\Psi.\cr
&&
\label{eq:THETATILDE}
\eea
$\varsigma,\tilde\xi, \tilde\rho, \tilde\omega$ have dimension $mass^{-2}$.

Notice that $\tilde\Phi$ involves the pseudoscalar singlet
$\ol\Psi\gamma_5 {\Bbb I}\Psi$.

We have now eight 4-plets of scalar fields, which exhausts the expected number
of particles transforming like a quark-antiquark pair.

\subsection{The quadratic invariant.}

The question of finding the quadratic invariants by ${\cal G}_s$ for the
scalar multiplets is important, since it will determine the form of the
possible gauge invariant kinetic and mass terms in the Lagrangian.

It is easy to show that for all the above 4-plets, the quadratic form is
unique and reads, in the basis corresponding to the charge eigenstates
operators (generators $t^+$ and $t^-$ instead of $t^1$ and $t^2$)
\be
{\cal Q}_4 = \left(\ba{rrrr}
                       1 & 0 & 0 & 0 \cr
                       0 & -1 &0 & 0 \cr
                       0 &0 & 0 & -1 \cr
                       0 & 0 & -1 &0
\ea\right);
\label{eq:Q4}\ee
the '$-$' signs are there because our pseudoscalars are purely imaginary.

This ends our formal considerations about the group of symmetry of the
Standard Model. We now study their dynamical consequences and the quantization
of the theory.
\section{The basic Lagrangian.}

I will show in section~\ref{sec:LINK} below that the composite
operators occurring in the above 4-plets scalar representations of
${\cal G}_s$ are the daily observed pseudoscalar and scalar mesons.
This will be done by studying in particular their leptonic decays and by
showing that they are in agreement with the usual `PCAC' computation.
The reader should not however be surprised if, already now, I speak of `pion'
or `kaon' \ldots, terminology which will be justified soon.

I also postpone the study of expressing the compositeness of the $\Phi$
multiplet, and its non-trivial consequences on the fermion spectrum in
particular. It will be done when dealing with quantum effects in
section~\ref{sec:QUANT}. For the moment, the reader can consider that writing
the scalar multiplets in terms of fermions was only a convenient way to study
their transformation by the gauge group.

The sole existence of the above scalar representations carries important
consequences; we can now build a gauge theory for the thirty-two scalars and
pseudoscalars present in the $N=4$ case. It is not our goal here to exhibit
all couplings but to emphasize some essential points.

In particular, we have now, in addition to $v$, the vacuum expectation value
of the Higgs boson, which controls the mass of the gauge fields and is
equivalent to $\mu$, {\em six} mass scales at our disposal, which will appear
in the Lagrangian with the corresponding quadratic invariants:
$M_\Phi$, $M_\Xi$, $M_\Delta$, $\tilde M_\Phi$, $\tilde M_\Xi$,
$\tilde M_\Delta$.
There is only one mass scale  associated to $\Delta$ and $\Theta$ because
they are related by complex conjugation and a real mass term will involve both.
The same occurs for $\tilde\Delta$ and $\tilde\Theta$.
We have consequently, in addition to $v$, two more mass parameters than the
number of `quark masses' in the QCD Lagrangian.

\subsection{The kinetic terms.}

They write
\bea
{\cal L}_S^{kin}&=&{1\over 2}\left(
{D_\mu}\Phi{\cal Q}_4 {D^\mu}\Phi^t
 + \kappa_1\, {D_\mu \Xi} {\cal Q}_4 {D^\mu \Xi}^t
      + \kappa_2\, {D_\mu \Delta} {\cal Q}_4 {D^\mu \Theta}^t \right.\cr
&+&\left. \kappa_3\, {D_\mu \tilde\Phi} {\cal Q}_4 {D^\mu \tilde\Phi}^t
+ \kappa_4\, {D_\mu \tilde \Xi} {\cal Q}_4 {D^\mu \tilde\Xi}^t
+ \kappa_5\, {D_\mu \tilde\Delta} {\cal Q}_4 {D^\mu \tilde\Theta}^t
\right).
\label{eq:LSKIN1}\eea
The superscript `$t$' means `transposed', as the scalar multiplets have been
written as line-vectors.
The $\kappa$ constants we choose so as to get the same normalization
as that for $\Phi$, yielding finally the very simple form
\be
{\cal L}_S^{kin} =
{1\over 2}{v^2\over N^2\mu^6} \sum_{(q_i=u,c,d,s)}\sum_{(q_j=u,c,d,s)}
D_\mu[\bar q_i q_j]D^\mu[\bar q_j q_i]-
D_\mu[\bar q_i\gamma_5 q_j]D^\mu[\bar q_j \gamma_5 q_i],
\label{eq:LSKIN2}\ee
where the notation ${v\over N\mu^3}[\bar q_i q_j]$ means the scalar field
transforming like this composite operator (same thing with the pseudoscalar
when a $\gamma_5$ is present).

\subsection{The mass eigenstates.}

In the same way, we write $SU(2)_L\times U(1)$ invariant mass terms:
\be
{\cal L}_S^{mass} =
-{1\over 2}(
            M_{\Phi}^2\ \Phi {\cal Q}_4 \Phi^t
        +   M_{\Xi}^2\ \Xi {\cal Q}_4 \Xi^t
	+   M_{\Delta}^2 \Delta {\cal Q}_4 \Theta^t
        +   \tilde M_{\Phi}^2\ \tilde\Phi {\cal Q}_4 \tilde\Phi^t
        +   \tilde M_{\Xi}^2\ \tilde \Xi {\cal Q}_4 \tilde \Xi^t
	+   \tilde M_{\Delta}^2 \tilde\Delta {\cal Q}_4 \tilde \Theta^t
           ).
\label{eq:LSM}\ee
Several remarks are in order:\l
- if one takes the same normalizations as above for the kinetic terms, the
mass terms also become `diagonal in the strong eigenstates' $[\bar q_i q_j]$
and $[\bar q_i\gamma_5 q_j]$ (pion, kaon \ldots), which then become degenerate
in $mass^2$; in general, this degeneracy is lifted, and the electroweak
eigenstates are not aligned with the strong (observed) ones; they are rather
generalizations of the $K^0_1, K^0_2 \ldots$ mesons (see for example
\cite{Lee}), involving more complicated combinations of the `constituent'
fermions;\l
- of course, the spectrum is also modified by the quartic terms that
one is free to --- and one has to --- add to the scalar potential,
compatible with the symmetry and with renormalizability
(cubic ones  should involve a $SU(2)$ singlet); they
generate, when the Higgs gets its vacuum expectation value, additional mass
terms which are in general not invariant by the gauge group, and which,
consequently, will again modify the alignment between strong and electroweak
eigenstates; relations between the different masses are expected to occur
if the number of independent scales happens to be lower than the number of
eigenstates;\l
- the multiplets mix scalars and pseudoscalars; unlike in the standard
picture of chiral symmetry breaking, we consequently expect mass
degeneracies between bound states of opposite parity, and not only between
pseudoscalars; a specially interesting case is $\tilde\Phi$, which involves
the pseudoscalar singlet and the scalar equivalents of the three $\varphi$'s
of the standard multiplet $\Phi$: they get their own mass scale
while no mass degeneracy has to be expected between the former and other
pseudoscalars; its having a heavier mass no longer  appears as a problem
($\eta'$); note that experimentally, the
mass degeneracy of the $\eta'$ with scalar mesons involving the u,d and s
quarks is well verified.

A detailed study of this spectrum  beyond the scope of this paper and is
postponed to a further work (see ref.~{Machet1}).
I hope that the reader already agrees that this
model provides something new with respect to the traditional gauge model
for quarks, with masses put by hand; explicit gauge invariance is
automatically achieved at the mesonic level and physical parameters are now
attached to experimentally observed asymptotic particles, not to `confined'
fields.

\subsection{The `Goldstone' particles.}

First, in the presence of the above invariant mass terms, the $2N^2 -3$
currents of the chiral group which do not correspond to the gauge currents
are not conserved;  it is for example simple to show that the only
invariance of the $\tilde\Phi {\cal Q}_4 \Phi^t$ quadratic term is by ${\cal
G}_s$; all other variations, non-vanishing, therefore correspond to
an explicit breaking not associated with any Goldstone-like particle.

The spontaneous breaking of the gauge group down to the $U(1)$ of
electromagnetism is, at the opposite, expected to give rise to three
Goldstone particles, the three $\varphi$'s. I however introduced an explicit
mass term for $\Phi$ and did not specially mention a `mexican hat'
potential, usually put at the origin of the breaking of the symmetry; the
$\varphi$'s are thus massive at the classical level (with mass $m_\pi$ as shown
in section~\ref{sec:TECHNI}), and not {\em stricto sensu} Goldstone
(massless) particles: this terminology is abusive here.

Next, I already mentioned that we would not look for the origin of the
mechanism giving rise to $\la \ol\Psi\Psi\ra = N\mu^3$; however, a suggestion
is implicit in section~\ref{sec:QUANT} below, dealing with quantum
effects. There, expressing the compositeness of $\Phi$ results in an
additional effective Lagrangian\l
- involving 4-fermions couplings;\l
- screening the classical scalar potential for $\Phi$.\l
It can be suggested that, in analogy with the work of Nambu and Jona-Lasinio
\cite{NambuJonaLasinio}, those couplings are at the origin of `quark
condensation' and of the dynamical breaking of chiral and gauge symmetries,
which are here the same phenomenon. More comments are made in
section~\ref{sec:NJL} below.

Now, among the $N^2$ pseudoscalar expected to be `light' in the standard
picture of chiral symmetry breaking,
none satisfies here such a criterion, and only {\em three} linear
combinations of the latter are `eaten' by the massive
gauge fields (see the general demonstration of unitarity in
section~\ref{sec:UNITARITY}). A conceptual problem associated with the spectrum
of pseudoscalar mesons has thus disappeared. The questions ``Why is the pion so
light?'' or ``Why is the $\eta'$ so heavy?'' are more easily answered
because the electroweak theory of mesons is able to
accommodate several mass scales. The link between the mass of the `pion' and
that of the gauge field will be studied in more detail below
(section~\ref{sec:TECHNI}), in relation
with `PCAC' and technicolour theories. The question of the scalar mesons
was still more difficult to apprehend, and they did not seem to fit in any
well established framework or classification. We have been able here to give
them a precise status, and close relationships are expected between them and
the former.

I naturally do not pretend that all conceptual problems have disappeared. In
particular, I will comment a little more, later, on the different status given
to three of the $2N^2$ `bound states'. This just means probably that the
present
tentative will have to be extended in the future towards a still more unified
framework.

{\em Remark:} only if the gauge group is $U(1)$ is there identity between
the (unique) pseudoscalar meson and the (unique) `Goldstone' boson.

\subsection{The link with observed particles.}\label{sec:LINK}

At this point, the Lagrangian is that of the Standard Model, ${\cal L}_{GSW}$,
 \cite{AbersLee}
to which has only been added the kinetic term for the additional scalar
multiplets ${\cal L}_S^{kin}$ and the associated mass terms ${\cal
L}_S^{mass}$. The other modifications are only conceptual.

Before going further in the reshuffling and reinterpretation of the model,
I will make the link with the usual pseudoscalar (and scalar) mesons,
showing that they are those introduced above.
This will proceed trough the
study of their leptonic decays. I will show that they have the same rates as
usually computed from `PCAC'. The proof will be completed in
section~\ref{sec:QUANT}, where I show that the `anomalous' decays of the
pseudoscalars
into two gauge fields are also recovered. As a consequence, invoking a new
scale of interactions, like in `technicolour' theories, is unneeded.

\subsubsection{The abelian example.}\label{sec:TOYMODEL}

Now, the intricacies of a non-abelian gauge group could only bring
useless technical complications to the demonstration. This is why I will
make the link with observed particles in the simplified case of a $U(1)_L$
gauge model. I hope that the reader will not be confused by having an
isosinglet pion instead of the real isotriplet, and will easily generalize to
the Glashow-Salam-Weinberg (GSW) Standard Model. A study of the non-abelian
case in the case of two generations of fermions is performed in
ref.~\cite{Machet1}.

Let us consider the generator ${\Bbb T}_L$ as the $4\times 4$ matrix
\be
{\Bbb T}_L = {1-\gamma_5\over 2}\; {\Bbb T},
\ee
and choose $\Bbb T$ to satisfy the condition
\be {\Bbb T}^2 = {\Bbb I},
\ee
which is the simplest case when the gauge generator $\Bbb T$ and the unit
matrix form an algebra.

I will deal in the following with the simple case where ${\Bbb T}$ is
itself the unit matrix
\be {\Bbb T}= {\Bbb I},\ee
but other cases can be considered as well.

The standard scalar multiplet reduces now to
\be
\Phi =(H,\varphi) = {v\over N\mu^3}
(\ol\Psi {\Bbb I}\Psi, -i\ol\Psi\gamma_5{\Bbb T}\Psi)
\label{eq:PHIAB}\ee
The group acts as follows on $\Phi$
\be\left\{\ba{lcl}
                {\Bbb T}_L.\varphi &=& iH,\cr
                {\Bbb T}_L.H    &=& -i\varphi. \ea \right .
\label{eq:GRACT}\ee
$H$ is written as usual
\be
H = v + h.
\ee
This case is so simple that the only other possible scalar doublet is that
obtained by a $\gamma_5$ transformation on $\Phi$, and is the same as
$\Phi$ itself. So, the Lagrangian writes
\bea
{\cal L} + {\cal L}_\ell &=& -{1\over 4}F_{\mu\nu}F^{\mu\nu} \cr
&+& i \ol\Psi\gamma^\mu (\p_\mu -ig\;\sigma_\mu \, {\Bbb T}_L )\Psi
+ i \ol\Psi_\ell\gamma^\mu (\p_\mu -ig\;\sigma_\mu \, {\Bbb T}_L )\Psi_\ell\cr
&+&{1\over 2}(D_\mu H D^\mu H + D_\mu\varphi D^\mu \varphi) -V(H^2 +
\varphi^2),
\label{eq:LAB}\eea
where we have introduced the coupling of the gauge field $\sigma_\mu$ to the
leptons $\Psi_\ell$. $V$ is the scalar potential.
One has:
\be\ba{lclcl}
D_\mu H &=& \p_\mu H -ig\sigma_\mu {\Bbb T}_L.H
                       &=& \p_\mu H -g\sigma_\mu\varphi, \\
D_\mu\varphi &=& \p_\mu\varphi -ig\sigma_\mu {\Bbb T}_L.\varphi
                       &=& \p_\mu\varphi +g\sigma_\mu H.
\ea\ee

The kinetic term for $\varphi$ involves the coupling
\be
gv\,\sigma_\mu\p^\mu\varphi;
\label{eq:NONDIAG}
\ee
 I show now that this triggers, through the
coupling of the gauge field to leptons, a coupling between $\varphi$ and the
leptons described by fig.~1, which yields the usual `PCAC' decay rate.
\figskip
\hskip 4cm\epsffile{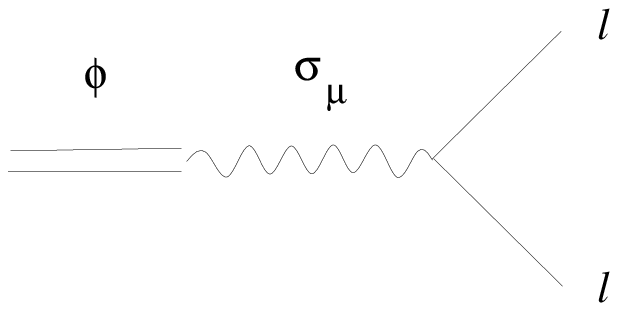}

{\centerline{\em Fig.~1: diagram generating the leptonic decays of $\varphi$.}}
\figskip
In the low energy regime, precisely relevant in the `PCAC' computations, the
propagator of $\sigma_\mu$ is $ig_{\mu\nu}/M_\sigma^2$ ($M_\sigma = gv$ is the
mass of the gauge field), such that fig.~1 gives the coupling
\be
-{i\over v} \p_\mu\varphi \ol\Psi_\ell \gamma^\mu {\Bbb T}_L \Psi.
\label{eq:PHILEPTONS}\ee
We now rescale the fields by
\bea
\varphi &=& a\pi,\cr
H &=& a H',\cr
\Psi &=& a\Psi',\cr
\Psi_\ell &=& a\Psi'_\ell,\cr
\sigma_\mu &=& a\, a_\mu,\cr
g &=& e/a.
\label{eq:SCALE}\eea

After a global  rescaling by $1/a^2$, the Lagrangian eq.~(\ref{eq:LAB})
 rewrites (we do not mention any longer the scalar potential)
\be\ba{lcl}
{\displaystyle {1\over a^2}}({\cal L}+{\cal L}_\ell) &=&
-{1\over 4}f_{\mu\nu}f^{\mu\nu}\cr
& & + i\,\ol\Psi' \gamma^\mu(\p_\mu -ie\,a_\mu {\Bbb T}_L)\Psi'
+i\,\ol\Psi'_\ell \gamma^\mu(\p_\mu -ie\,a_\mu {\Bbb T}_L)\Psi'_\ell\cr
& & +{1\over 2}\Big((\p_\mu H' -e \,a_\mu \pi)^2 +
(\p_\mu \pi +e\, a_\mu H')^2\Big)
\label{eq:LSCALED}\ea\ee
where
\be
f_{\mu\nu} = \p_\mu a_\nu - \p_\nu a_\mu.
\ee
We have
\be
\la H'\ra = {v\over a},\ \la\ol\Psi' \Psi'\ra = {N\mu^3\over a^2},
\ee
and
\be
e^2 \la H'\ra ^2 = g^2 \la H \ra ^2,
\ee
yielding the same mass $M_\sigma$ for $a_\mu$ and $\sigma_\mu$.
 We call  now $a_\mu$ `vector boson', $\pi$ `pion',
identify the `primed' leptons with the observed ones and $e$ with the
`physical' coupling constant of the theory.
Then, from $({\cal L} +{\cal L}_\ell)/a^2$, we get an effective coupling,
equivalent to eq.~(\ref{eq:PHILEPTONS})
\be
-i\,{a\over  v} \;
\ol\Psi'_\ell \gamma_\mu {\Bbb T}_L \Psi'_\ell\;  \p^\mu \pi
\label{eq:PILEPTONS}\ee
which rebuilds the correct S-matrix element, as computed traditionally by
`PCAC', for the decay of the pion into two leptons, when one takes
\be
a = {f_\pi\over v}.
\label{eq:A}\ee
This shows that $\varphi$ decays into leptons like an isoscalar pion, and
that there is no contradiction between
dynamically breaking  the gauge symmetry \cite{FarhiJackiw}
 and identifying the longitudinal
component of the massive gauge field as the usual pseudoscalar meson.

\subsubsection{The rescaling of the fields.}\label{sec:SCALING}

The rescaling eq.~(\ref{eq:SCALE}) clearly deserves more comments.
When we consider the
rescaled Lagrangian ${\cal L}/a^2$, the kinetic terms of all new (primed)
fields stay normalized to $1$, and all masses (poles of the bare
propagators) are left unchanged. One easily sees that the new
Lagrangian has the same expression in terms of the new fields and coupling
($e$) as the original one in terms of the original fields and $g$,
except for the terms quartic in the scalars (which I did not write explicitly)
or those appearing in the Lagrangian of constraint. However, in the latter,
the limit $\beta\rar 0$ makes things stay qualitatively unchanged at the
classical level.
The (non-perturbative) wave function renormalization can modify effective
couplings, either appearing
at tree level, like the coupling between mesons and leptons studied above, or
 through loops, like the `anomalous' couplings of pseudoscalars to two gauge
fields to be studied in sections~\ref{sec:ABANO} and \ref{sec:NABANO}.
One has indeed to be specially careful when computing quantum effects,
as the generating functional now involves
\be
e^{i{a^2\over h\!\!\!/}\int d^4x{\cal L}(x)/a^2},
\label{eq:LOOP}\ee
where ${\cal L}/a^2$ is the Lagrangian used at the classical level, expressed
in terms of the rescaled fields. As can be read in eq.~(\ref{eq:LOOP}),
the parameter controlling the loop expansion \cite{Coleman73}, instead of
being the Planck's constant $h\!\!\!/$, has become $h\!\!\!//a^2$.

These points are further developed in ref.~\cite{Machet1}.

\subsubsection{A remark on gauge fixing.}

The gauge fixing in massive gauge theories is usually chosen \cite{AbersLee}
so as to cancel
the non-diagonal coupling eq.~(\ref{eq:NONDIAG})
  between the gauge fields ($\sigma_\mu$)
and the would-be Goldstone bosons ($\varphi$); `gauging away' this coupling
gauges the leptonic decays of $\varphi$ into that of the longitudinal
component of the massive $\sigma_\mu$: there is indeed identity between those
two states: the `pion' {\em is} the third component of $\sigma_\mu$. I
will explicitly demonstrate unitarity in the next section.

\subsubsection{A comment about previous works.}\label{sec:PREVIOUS}

In  a previous work \cite{Machet2}, I have introduced, according to
\cite{BabelonSchaposnikViallet}, a derivative coupling between a
Wess-Zumino field $\xi$ \cite{WessZumino}, closely related with $\varphi$,
and the gauge
current.  In its presence, the PCAC equation linking the divergence of the
gauge current with the pseudoscalar field could be deduced from the
equations for $\varphi$, but its contribution was shown to be canceled
as a consequence of the coupling eq.~(\ref{eq:NONDIAG}) above.
Finally, we recovered the correct leptonic decay
of the pion from this derivative coupling itself, and the result was
globally the same. I now prefer {\em not to introduce this
coupling}, which can only complicate the renormalization of the theory.
It might be innocent because the gauge current turns out to be exactly
conserved (at the operator level), such that it could be legitimate to take
its divergence identically vanishing in the Lagrangian. However, the
subtleties associated with this kind of manipulation can always give rise to
criticism. It is reassuring that we find the same physical results, which
thus do not depend on the introduction or not of such a term.

The motivation, in \cite{BabelonSchaposnikViallet}, for introducing it
coupling, was the recovery of gauge invariance in a gauge theory with
anomalies:
the corresponding gauge transformation acted on $\sigma_\mu$ and on the
Wess-Zumino field $\xi$ (see also section~\ref{sec:QUANT}), such that
the combination $\sigma_\mu -(1/g) \p_\mu\xi$ was invariant.
The effective action occurred to be only a function of this
combination, hence the invariance. It is however no longer assured here when
we introduce, in the Feynman path integral, (see section~\ref{sec:QUANT})
 the constraints expressing the compositeness of $\Phi$, not
invariant by such a transformation.
Furthermore, as we shall see, the present theory becomes anomaly-free by
another mechanism. The corresponding gauge invariance is thus not the one
evoked above, but the usual one acting on the fermions, the scalars and the
gauge field, and it remains `unspoiled'.

\subsubsection{A comment about `PCAC' and technicolour theories.}
\label{sec:TECHNI}

In the `technicolour' framework of dynamical symmetry breaking
\cite{SusskindWeinberg}, where two
massless poles, that of a gauge field and that of a Goldstone boson
`transmute' into that of a massive gauge field, there is a mismatch by the
factor $a = {f_\pi / v}$ eq.~(\ref{eq:A}) between the mass of the gauge field
and that of the observed $W$'s when the Goldstone is identified with the pion.
Or, if one fits the $W$ mass, one is forced to introduce new
ultra-heavy `technipions' for the scheme to be coherent.
A simple argument shows how this problem has been cured.
We recall  the usual PCAC statement linking the pion with the divergence of the
corresponding axial current $J_\mu^5$
\be
\p^\mu J_\mu^5 = i\,f_\pi\, m_\pi^2\, \pi.
\label{eq:PCAC1}
\ee
Eq.~(\ref{eq:PCAC1}) stays qualitatively unchanged when deduced from the group
invariant mass terms introduced for the scalar multiplets. For this purpose,
we go back to the non-abelian case since, in the $U(1)$ case, the chiral
current is identical with the gauge current and is exactly conserved.
There, and when the mixing angle is non vanishing, one
sees immediately that, for example the axial current  $J^{1+i2}_{\mu\, 5} =
\bar u \gamma_\mu\gamma_5 d$, with the quantum number of the charged pion
$\pi^+$, is not conserved: by varying the Lagrangian with a global axial
transformation carrying this quantum number, one finds, from the mass term
$-1/2\, M_\Phi^2 \Phi^2$, the divergence:
\be
\p^\mu J^{1+i2}_{\mu\, 5} =
{i\over 2} M_\Phi^2 ({2v^2\over N\mu^3}\bar u \gamma_5 d
                                           -v\cos\theta_c\;\phi^+) + \cdots
                          = i\,\alpha\, M_\Phi^2\,  v\, \varphi^+ +\cdots
                          = i\,\alpha'\, M_\Phi^2\, f_\pi\ \pi^+ +\cdots,
\label{eq:PCAC2}\ee
where we have used eq.~(\ref{eq:SCALE}) to relate $\varphi$ and $\pi$,
 $\alpha$, $\alpha'$ are numerical coefficients of order $1$, and
the `$\cdots$' involve terms proportional to $\bar c\gamma_5 s$,
corrections in $\sin\theta_c$, terms quadratic in the $\varphi$'s
and contributions from other terms in ${\cal L}_s^{mass}$.
This is eq.~(\ref{eq:PCAC1}), with
\be
M_\Phi = m_\pi.
\label{eq:MPHIPI}
\ee
The important point to notice is that $v$ takes the place of $f_\pi$ in the
PCAC equation written with the original fields, which is exactly what is
needed to make technicolour `work'.
Going to the `primed' leptonic fields by the equivalent of
eq.~(\ref{eq:SCALE}) we get now
\be
\p^\mu J^{'1+i2}_{\mu\, 5} = \alpha'\, m_\pi^2\,
{v^2\over f_\pi^2}\, f_\pi\, \pi^+ +\cdots.
\label{eq:PCAC3}\ee
If we perform a `traditional' PCAC computation of the leptonic decay of
$\pi$, using eq.~(\ref{eq:PCAC3}) {\em and with  the rescaled charge
$e = (f_\pi/v)g$}, we recover the correct result.
This shows that the introduction of a technicolour scale is unneeded.

It is also evident that our model is free from flavour changing neutral
currents, and, consequently, that this other crucial problem of technicolour
theories has also found a natural solution.

\section{Quantizing.}\label{sec:QUANT}

I dealt above with some classical aspects of the model. I will now study
its quantization and some consequences of this process. As  the least action
principle is the basic unifying principle of all physics, I will use the
Feynman path integral method \cite{Feynman,Popov}.

All results of this section are the consequences of one remark:
by identifying the gauge group as a subgroup of the chiral group and the two
phenomena of gauge and chiral symmetry breaking, we have {\em explicitly}
considered the Higgs boson, and its three companions in the basic scalar
4-plet $\Phi$, as composite fields. However, when performing the path
integration, one traditionally integrates on both the quarks and the four
components of $\Phi$, that is, now, on non-independent degrees of freedom.
The consistency of this approach can only be achieved if constraints are
introduced in the path integral, in the form of $\delta$-functionals,
explicitly relating the above multiplet to its component fermions.
Those constraints  I will write into an exponential form, equivalent to
introducing an effective additional Lagrangian ${\cal L}_c$. It involves
4-fermions couplings which I shall study in the leading approximation in a
development in inverse powers of $N$, the number of flavours, corresponding
to the approximation of Nambu and Jona-Lasinio \cite{NambuJonaLasinio}.
 In this context I show that
they turn out to be renormalizable, transmuting into a vanishing effective
coupling when the appropriate resummation is performed. Those 4-fermions
couplings, of course, also affect the quark mass and, in reality, both
 satisfy a system of two coupled equations.
When the Higgs boson gets its non-vanishing vacuum expectation value $v$, a
bare infinite quark mass springs out from ${\cal L}_c$; this solution is
shown to be stable by the above system of equations. The quarks consequently
become unobservable, because infinitely massive, and their degrees of
freedom have been transmuted into those of the scalars: the fermionic {\em
fields} of the Lagrangian do not correspond to {\em particles}. This is in
agreement with  the property of the Nambu Jona-Lasinio approximation to
propagate only bound states, and shows the consistency of our approach.
A theory with infinitely massive fermions becomes anomaly-free, as can be
easily shown in the Pauli-Villars regularization \cite{PauliVillars}
 of the `triangle diagram',
which yields the covariant form of the anomaly. This is to be related with a
previous study of ours \cite{BellonMachet},
 showing that the electroweak interactions of leptons
could be considered as coming from a purely vectorial theory, thus also
anomaly-free. The two sectors of quarks and leptons, being now
both and independently anomaly-free, can be safely disconnected. Indeed,
only the cancelation of anomalies imposed that the number of leptons should
match that of the quarks \cite{BouchiatIliopoulosMeyer}, linking together
two totally different types of objects.

The reader will probably feel uneasy with the other $2N^2-4$ scalar and
pseudoscalar particles which we also formally introduced as composite
operators. We were careful to mention that this was only to easier apprehend
their symmetry properties and allow a more intuitive treatment.
Indeed, at no point do I introduce constraints for those states, unlike
for the standard multiplet $\Phi$. Does this means that there is no
necessity to explicitly consider them as quark-antiquark bound states? the
following answers can only reflect my own prejudice:\l
- as it is, the Standard Model only couples the massive $W$ gauge fields to
some precise quark combinations, or, phrased in another way, the
gauge currents have very particular forms: they are controlled by the Cabibbo
angle and, for example, they do not involve any flavour changing
neutral current. It is `natural' to make a link between those massive
gauge fields and composite objects, and to consider them `made of' the
combinations of the quarks that they are coupled to (a massless gauge field
stays conceptually a fundamental object). Then, the longitudinal components of
the massive objects
are themselves naturally composite; but they are precisely the three
companion of the Higgs boson in $\Phi$. The other scalars and pseudoscalars
bearing no relation whatsoever with massive composite gauge fields,
nothing forces them to be themselves composite;\l
- the fact that the Higgs boson is of the form $\ol\Psi\Psi$, unifying
the pictures of gauge and chiral breaking, leads, for consistency, its three
companions to be also explicitly composite. The other scalar multiplets do
not involve $H$ and thus are not subject to this conceptual constraint;\l
- from a technical point of view, introducing constraints for the other
multiplets appears inconsistent: their exponential form tends to decouple
the associated scalars and pseudoscalars themselves by giving them infinite
masses, which is physically not welcome.

Thus, in its present state, the Standard Model will be considered to admit
{\em only four} composite eigenstates which are the components of the standard
multiplet $\Phi$. The pressure is of course high for a higher level of
unification, where all scalars and pseudoscalar `strong' eigenstates would
recover a composite picture that we are accustomed to work with;
in my opinion, this could only be achieved in a $U(N)_L\times U(N)_R$ gauge
theory, that is if the gauge group {\em is} the chiral group. Then and only
then could we also claim to have a unification of the three types of
interactions.

\subsection{Quantizing with non-independent degrees of freedom.}

I shall work again here with the abelian model used in
section~\ref{sec:TOYMODEL}. All basic ingredients and results are present
but not the useless intricacies due to the non-abelian group.

We introduce in the path integral  $\delta$-functionals expressing the
compositeness of $H$ and $\varphi$:
\be
\prod_x \delta(C_H(x))\prod _x \delta (C_\varphi(x)),
\ee
with
\be
C_H= H-{v\over N\mu^3}\ol\Psi\Psi,
\label{eq:CH}\ee
\be
C_\varphi=\varphi +i\,{v\over N\mu^3}\ol\Psi\gamma_5{\Bbb T}\Psi,
\label{eq:Cphi}\ee
and define the theory by
\be Z= \int {\cal D} \Psi{ \cal D} \ol\Psi{\cal D}H
{\cal D} \varphi {\cal D} \sigma_\mu \  e^{i\int d^4 x {\cal L}(x)}
\prod _x \delta(C_H(x))\prod _x \delta (C_\varphi(x)).\label{eq:Z} \ee
Rewriting the $\delta$ functionals in their exponential form, we transform
them into the effective Lagrangian ${\cal L}_c$, that I will call
`Lagrangian of constraint':
\be
{\cal L}_c =
\lim_{\beta\rar 0}\frac{-N\Lambda^2}{2\beta}
\left(H^2 +\varphi^2 -{2v\over N\mu^3}(H\ol\Psi\Psi-i\,\varphi
\ol\Psi\gamma_5 {\Bbb T} \Psi)+
 {v^2\over N^2\mu^6}\left((\ol\Psi\Psi)^2
-(\ol\Psi\gamma_5 {\Bbb T} \Psi)^2\right)\right).
\label{eq:LCAB}\ee
$\Lambda$ is an arbitrary mass scale.\l
{\em Remark 1}: we have exponentiated the two constraints on $H$ and $\varphi$
with the same coefficient $\beta$: it makes ${\cal L}_c$ gauge invariant
(the gauge transformation acting on both scalars and fermions) and eases the
computations.\l
{\em Remark 2}: we see clearly on eq.~(\ref{eq:LCAB}) that, after
integrating over the fermions, the effective Lagrangian cannot have
trivially the gauge invariance of ref.~\cite{BabelonSchaposnikViallet}, but the
usual one acting on the gauge field, the fermions and the scalars.
See also section~\ref{sec:PREVIOUS}.

\subsubsection{Unitarity.}\label{sec:UNITARITY}

It is more easily studied by going from $H,\varphi$ to the variables $\tilde
H,\xi$, both real, defined by (see \cite{AbersLee})
\be \tilde H =e^{-i\frac{\xi}{v}{\Bbb T}_L}.\
(H+i\varphi),\label{eq:CHVAR1}\ee
with
\be \tilde H = v+\eta.\label{eq:shift2}\ee
The solution of eq.~(\ref{eq:CHVAR1}) is
\be \left\{\ba{rcl}
0 &=& H\sin{\xi\over v}+\varphi\cos{\xi\over v}, \\
\tilde H &=& H\cos{\xi\over v}-\varphi\sin{\xi\over v},
\ea\right .\label{eq:CHVAR2}\ee
from which $\eta$ and $\xi$ can be expressed as series in $h/v$ and
$\varphi/v$:
\be\ba{lcl}
\xi &=& -\varphi\,(1-{h\over v} + {h^2\over v^2}- {\varphi^2 \over 3v^2} +
\cdots\ ),\\
\eta &=& h+{\varphi^2\over 2v}(1-{h\over v}) +\cdots .
\ea\label{eq:CHVAR3}\ee
The laws of transformation of $\tilde H$ and $\xi$ come from
eqs.~(\ref{eq:GRACT}) and (\ref{eq:CHVAR2}):
\be
when\qquad \Psi \lrar e^{-i\theta {\Bbb T}_L} \Psi,
\ee
\be\left\{\ba{lcl}
\xi &\lrar & \xi -\theta v,\\
\tilde{H} &=& invariant.
\ea\right .\label{eq:TRANS}\ee
A gauge transformation induces a translation on the field $\xi$,
equivalent to:
\be
e^{i{\xi\over v}{\Bbb T}_L} \lrar e^ {-i\theta {\Bbb T}_L}\
e^ {i{\xi\over v}{\Bbb T}_L}.
\ee
Eq.~(\ref{eq:TRANS}) corresponds to a non-linear realization of the gauge
symmetry \cite{WessWeinbergZumino}.
$\xi$ is a natural Wess-Zumino field \cite{WessZumino}.

{}From the definition of $\tilde H$ and $\xi$ in eq.~(\ref{eq:CHVAR1}), we
have
\be
\ti H^2 = H^2 +\varphi^2,
\ee
and
\be
{1\over 2}(D_\mu H D^\mu H +D_\mu\varphi D^\mu\varphi)=
{1\over 2} \p_\mu \tilde H \p^\mu \tilde H +{1\over 2}g^2(\sigma_\mu
-{1\over g}\p_\mu{\xi\over v})^2\tilde H^2,
\label{eq:LS1}\ee
such that $\cal L$ also writes
\be\ba{ccl} {\cal L}(x)&=& -{1\over 4} F_{\mu \nu}F^{\mu\nu}
+i\ol\Psi \gamma^\mu
\left(\p_\mu -ig\sigma _\mu\,{\Bbb T}_L\right)\Psi\\ & &
+{ 1\over 2} \p_\mu \tilde H \p^\mu \tilde H
+{1\over 2} g^2\left(\sigma _\mu -(1/g)\ \p_\mu \xi/v\right)^2
\tilde H^2 - V(\tilde H^2). \ea\label{eq:L1}
\ee
We perform in $Z$ the change of variables
\be
\left\{\ba{rcl}
\Psi &\lrar & e^{-i(\chi/v){ \Bbb T}_L}\Psi,\\
H+i\varphi &\lrar & e^{-i(\chi/v) {\Bbb T}_L}.\ (H+i\varphi);
\ea\right .
\label{eq:CHVAR}\ee
it leaves ${\cal L}_c$ invariant and yields two Jacobians:\l
-  the first, coming from the transformation of the fermionic measure
\cite{Fujikawa79}, is
\be
J = e^{i\int d^4x {(\chi/v){\cal A}}},
\ee
where $\cal A$ is the (eventual) anomaly \cite{AdlerBellJackiw};\l
-  the second, corresponding to a `rotation' of the scalars, is unity.

 We use the laws of transformation
eq.~(\ref{eq:TRANS}) and the fact that the scalar Lagrangian ${\cal L}_s$
\be\ba{lcl}
{\cal L}_s &=& { 1\over 2} \p_\mu \tilde H \p^\mu \tilde H
+{1\over 2} g^2\left(\sigma _\mu -(1/g)\ \p_\mu \xi/v\right)^2
\tilde H^2 - V(\tilde H^2)\\
&=&
{1\over 2}(D_\mu H D^\mu H +D_\mu\varphi D^\mu\varphi) -V(H^2 + \varphi^2)
\ea\label{eq:LS2}\ee
is invariant when one transforms both the gauge field and the scalars:
\be
{\cal L}_s(\xi-\theta v, \tilde H,\sigma_\mu) =
{\cal L}_s\left(\xi, \tilde H, \sigma_\mu +(1/g)\,\p_\mu \theta\right),
\ee
to deduce that, by the change of variables eq.~(\ref{eq:CHVAR}), one gets an
effective Lagrangian
\be\ba{lcl}
{\cal L}'+ {\cal L}_c &=&
-{1\over 4}F_{\mu\nu}F^{\mu\nu} +i\ol\Psi
\gamma^\mu(\p_\mu -ig\sigma_\mu {\Bbb T}_L)\Psi\\
& &
+{1\over 2}\left(
\left(\p_\mu H -g(\sigma_\mu +{1\over g}{\p_\mu \chi\over
v})\varphi\right)^2
+\left(\p_\mu \varphi +g(\sigma_\mu +{1\over g}{\p_\mu \chi\over
v})H\right)^2
\right)
-V(H^2 +\varphi^2)\\
& & -(\chi/ v)\  (\p^\mu J_\mu^\psi -{\cal A}) \\
& & + {\cal L}_c.
\ea\label{eq:LPRIME}\ee
Some explanations are in order:\l
\quad - the $-(\chi/v)\,\p^\mu J_\mu^\psi$ term comes from the transformation
of the fermions;\l
\quad - the $(\chi/v)\,{\cal A}$ comes from $J$.

We then choose $\chi = \xi$ and finally go to the integration variables
$\xi$ and $\tilde H$ defined by eqs.~(\ref{eq:CHVAR1}) and
(\ref{eq:CHVAR2}).
This yields one more Jacobian $J_1$ according to
\be
{\cal D} H {\cal D} \phi = J_1\  {\cal D} \tilde H {\cal D}
\xi,
\ee
which can be expressed as
\be
J_1 =\prod _x {\ti H(x)\over v}=
\exp \left(\delta^4(0)\int d^4x\sum_{n=1}^{\infty}
{(-1)^{(n+1)}}{({\eta/v)^n}\over n}\right).
\label{eq:J1}\ee
Finally, after the two transformations above, Z becomes
\be
Z = \int {\cal D} \Psi{ \cal D} \ol\Psi
{\cal D} \tilde H {\cal D} \xi {\cal D} \sigma_\mu \  J_1 \
e^{i\int d^4 x \left(\ti{\cal L}(x)+ {\cal L}_c(x)\right)},
\label{eq:ZTILDE}\ee
with, using again eq.~(\ref{eq:LS1}),
\be\ba{ccl}
\ti{\cal L}& =&
-{1\over 4}F_{\mu\nu}F^{\mu\nu} +i\ol\Psi\gamma^\mu
\left( \p_\mu -ig\sigma_\mu {\Bbb T}_L\right)\Psi
+(\xi/v)\,{\cal A}-(\xi/v)\,\p^\mu J_\mu^\psi\\
& &
+{1\over 2}\p_\mu\tilde H\p^\mu\tilde H +{1\over 2}\sigma_\mu^2 \tilde H^2
-V(\tilde H^2).
\ea\label{eq:LTILDE}\ee
As $J_1$ in eq.~(\ref{eq:J1}) does not depend on $\xi$, the $\xi$ equation
coming from eqs.~(\ref{eq:ZTILDE}) and (\ref{eq:LTILDE}) is now
\be
\p^\mu J_\mu^\psi -{\cal A} - v{\p{\cal L}_c\over \p\xi}=0.
\label{eq:XIEQ}\ee
$v\ {\p{\cal L}_c /\p\xi}$ is the classical contribution
$\p_\mu J^\mu_{\psi c}$, coming from ${\cal L}_c$, to the divergence of the
fermionic current, as can be seen by varying the Lagrangian
${\cal L} + {\cal L}_c$ with a global fermionic transformation; indeed,
${\cal L}_c$ is invariant by a global as well as local $U(1)_L$
transformation, and $\ti H$ being itself invariant by construction, the
variation of ${\cal L}_c$ with respect to $\xi$ cancels that with respect to
the fermions. Such that eq.~(\ref{eq:XIEQ}) rewrites
\be
\p^\mu J_\mu^\psi -{\cal A} -\p^\mu J_{\mu c}^\psi =0,
\ee
and is nothing more than the exact  equation for $\p^\mu J_\mu^\psi$.
We have thus put $Z$ in a form where $\xi$ appears as a Lagrange multiplier;
the associated constraint being satisfied at all orders, $\xi$ disappears:
it was just an `auxiliary field', which could be
`gauged away' and transformed into the third polarization of the
massive vector boson.  This form of the theory describing a massive gauge
field is manifestly unitary. The transformations performed above
 are equivalent to going to the `unitary gauge' (see \cite{AbersLee}).

\subsection{Similarity and difference with the approach of Nambu and
Jona-Lasinio.}\label{sec:NJL}

The 4-fermions interactions in eq.~(\ref{eq:LCAB}) can be compared with
those introduced by Nambu and Jona-Lasinio \cite{NambuJonaLasinio} in their
approach of dynamical symmetry breaking. They differ in two respects,
extensively developed in the rest of the paper:\l
- the bare 4-fermions couplings are infinite when $\beta\rar 0$, and the
effective ones, obtained by resummation, go to $0$ in the same limit;\l
- the fermion masses are also infinite at this limit when $\la H\ra = v$.\l
The first difference leaves brighter prospects for renormalizability than in
the original paper; the second enables the disappearance of the anomaly and
the transmutation of the fermionic degrees of freedom into mesonic ones.

Despite those differences, the hope is that ${\cal L}_c$ can trigger
dynamical symmetry breaking by generating $\la\ol\Psi\Psi\ra \not = 0$. This is
under investigation. Proving that would shed some light on the mechanism
which underlines quark condensation, taken here as a given fact and a
constraint on the theory. It is usually attributed to strong interactions;
the demonstration of the above conjecture would be another hint (see below
section~\ref{sec:STRONG}) that the border between electroweak and strong
physics could be thinner than usually considered, and that the eventual
springing of strong interactions from the Standard Model of electroweak
interactions itself is worth investigating in details.

\subsection{The Nambu Jona-Lasinio approximation.}

Because of the potentially non-renormalizable 4-fermions couplings of
${\cal L}_c$, the theory defined by
\be
Z =  \int {\cal D} \Psi{ \cal D} \ol\Psi{\cal D}H
  {\cal D} \phi {\cal D} \sigma_\mu \  e^{i\int d^4 x ({\cal L} + {\cal
L}_c)(x)}
\ee
could be thought {\em a priori} pathological.
We shall however see that in the approximation of resumming `ladder diagrams'
of 1-loop fermionic bubbles or, equivalently, of dropping contributions at
order higher than $1$ in an expansion in powers of $1/N$ (Nambu Jona-Lasinio
approximation \cite{NambuJonaLasinio}), special properties are
exhibited:\l
\quad - the effective 4-fermions couplings go to $0$ and the effective
fermion mass goes to infinity;\l
\quad - the scalar $h$ (or $\eta$) also decouples.

In this approximation, our model will be shown to be a gauge invariant,
anomaly-free theory, the only asymptotic states of which are the three
polarizations of the massive gauge field (one of them being the composite
field $\varphi$, shown above  to behave like an abelian pion).
Isolated quarks are no longer observed as `particles', showing the
consistency of this approximation, known to propagate only fermionic bound
states \cite{NambuJonaLasinio}.

The analysis being based on truncating an expansion in powers of $1/N$,
we make precise our counting rules:\l
\quad - $g^2$ is of order $1/N$ (see \cite{Coquereaux});\l
\quad - the 4-fermions couplings are of order $1/N$ (see
eq.~(\ref{eq:LCAB}));\l
\quad - from the definitions of $h$ and $\varphi$ ($\eta$ and $\xi$), their
propagators are also of order $1/N$, a factor $N$ coming from the associated
fermionic loop;\l
\quad - thus, we shall consistently attribute a power $N^{-1/2}$ to the
fields $h, \varphi, \eta, \xi$, and $N^{1/2}$ to fermion bilinears
including a sum over the flavour index, such that, as expected,
$g\,\sigma^\mu\,\ol\Psi\gamma_\mu{\Bbb T}_L\Psi$ is of order $1$ like
$\sigma_\mu$ itself and the whole Lagrangian $\cal L$.

\subsubsection{The effective fermion mass and 4-fermions couplings.}

As can be seen in eq.~(\ref{eq:LCAB}), when $\la H\ra = v$,
${\cal L}_c$ introduces\l
\quad - an infinite bare fermion mass:
\be
m_0 = - {\Lambda^2 v^2\over \beta\mu^3};
\label{eq:M0}\ee
\quad - infinite 4-fermions couplings
\be
\zeta_0 = - \zeta_0^5 = {m_0\over 2N\mu^3}.
\label{eq:ZETA0}\ee
At the classical level, the infinite fermion mass in ${\cal L}_c$
is canceled by the 4-fermions term $\propto(\ol\Psi\Psi)^2$ when
$\la\ol\Psi\Psi\ra = N\mu^3$; however, staying in the `Nambu Jona-Lasinio
approximation' \cite{NambuJonaLasinio}, equivalent to keeping only diagrams
leading in an expansion in powers of $1/N$, the fermion mass and the
effective
4-fermions coupling satisfy the two coupled equations
\bea
\zeta(q^2) &=& \frac{\zeta_0}{1-\zeta_0 \; A(q^2, m)},\cr
\noalign{\vskip 2mm}
m &=& m_0 -2\zeta(0)\mu^3,
\label{eq:ZETAM}\eea
graphically depicted in fig.~2 and fig.~3. $A(q^2,m)$ is the one-loop
fermionic bubble. The above cancelation represents only the first two terms
of the series depicted in fig.~3.

\figskip
\epsffile{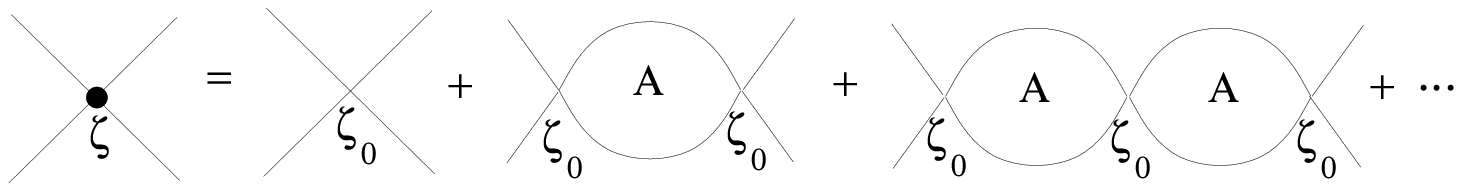}

{\centerline{\em Fig.~2: the effective 4-fermions coupling $\zeta(q^2)$.}}
\figskip
\hskip -.5cm\epsffile{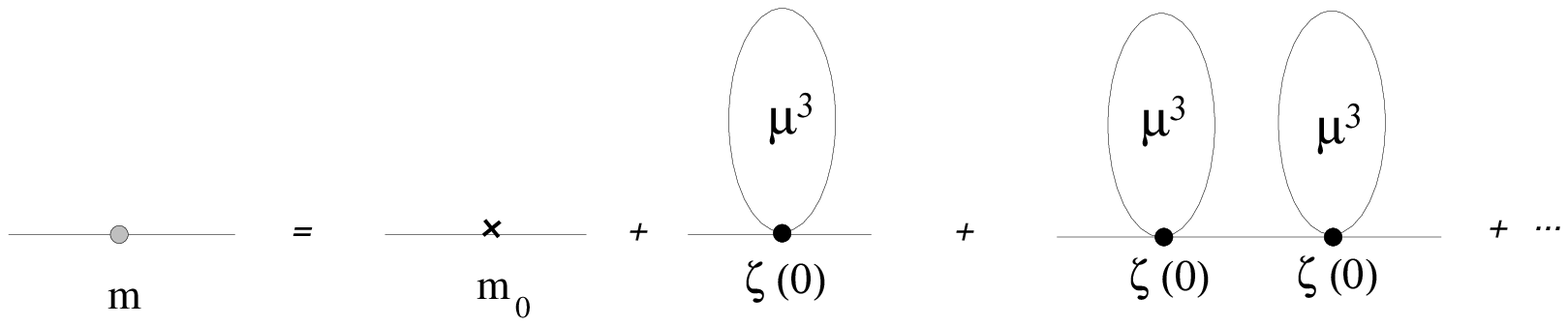}

{\centerline{\em Fig.~3: resumming the fermion propagator.}}
\figskip
$\mu^3$ being finite, $m=m_0$ is a solution of the equations
(\ref{eq:ZETAM}) above as soon as $\zeta(0)$ goes to $0$.
This is the case here since $\zeta(0)\propto -A(0,m)^{-1}$, and $A$ involves
a term proportional to $m^2$ (see for example \cite{Broadhurst}).
This also makes the effective 4-fermions coupling $\zeta(q^2)$
(and similarly $\zeta^5(q^2)$) go to $0$ like $\beta^2$.
The fermions are thus infinitely massive, which is exactly what is expected
for fields which do not appear as asymptotic states (particles). The `mass'
occurring here has of course nothing to do with the so-called `quark
masses', either present in the Lagrangian of Quantum Chromodynamics
(`current' masses) or in the Quark Model (`constituent' masses), which are
phenomenological parameters.

\subsubsection{The Higgs field.}

The scalar potential, usually chosen as
\be
V(H,\varphi)= -{\sigma^2\over 2}(H^2+\varphi^2)+
{\lambda\over 4}(H^2+\varphi^2)^2
\ee
is modified by the constraint in its exponentiated form, to become
\be\ba{lcl}
\hskip -0.5cm\ti V(H,\varphi)&=& V(H,\varphi)\\
&+&\lim_{\beta\rar 0}
{N\Lambda^2\over 2\beta}\left(H^2 +\varphi^2 -
{2v\over N\mu^3}(H\ol\Psi\Psi-i\,\varphi \ol\Psi\gamma_5
{\Bbb T} \Psi)
+ {v^2\over N^2\mu^6}\left((\ol\Psi\Psi)^2
-(\ol\Psi\gamma_5 {\Bbb T} \Psi)^2\right)\right).
\ea\label{eq:VTILDE}\ee
Its minimum still corresponds to $\langle H\rangle = v$,
$\langle \ol\Psi \Psi \rangle =N\mu^3$, $\langle \varphi\rangle
= 0$ if $\sigma^2 =\lambda v^2$, but the scalar mass squared has now become
\be
\left .\frac{\p^2\ti V}{\p H^2}\right|_{H=v}= -\sigma^2 + 3\lambda
v^2 +{N\Lambda^2\over\beta},
\ee
which goes to $\infty$ at the limit $\beta\rar 0$ when the constraints are
implemented.

The coupling between the scalar and the fermions present in ${\cal L}_c$
does not modify this result qualitatively. Indeed, resumming the series
depicted in fig.~4, the scalar propagator becomes
\be
D_h = {\displaystyle\frac{D^0_h} {1-D^0_h
  \left(\frac{i\Lambda^2 v}{\beta\mu^3}\right)^2 A(q^2,m)}}\label{eq:HPROPAG},
\ee
where $D^0_h$ is the bare scalar propagator
\be
D^0_h = \frac{i}{q^2 -{N\Lambda^2\over\beta}}.
\ee
At high $q^2$, $A(q^2,m)$ behaves like $N\,(b_1\,q^2 + b_2\,  m^2 + \ldots\,)$
(see for example \cite{Broadhurst}), such
that, when $\beta \rar 0$, $D_h$ now gets a pole at $q^2 = -(b_2/b_1)\,m^2$.
Checking \cite{Broadhurst} that the sign of $-b_2/b_1$ is positive confirms
the infinite value of the mass of the scalar. Furthermore, in this same
limit, $D_h$ goes to $0$ like $\beta^2$.

We conclude that the scalar field $h$ (nor, similarly, $\eta$)
cannot be produced either as an asymptotic state.
\figskip
\epsffile{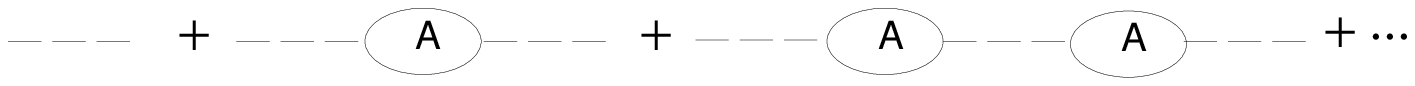}
\figskip

\centerline {\em Fig.~4: resumming the scalar propagator.}
\figskip
{\em Remark 1:} ${\cal L}_c$ screens the classical scalar potential for
$\Phi$ and one could expect that, before being `eaten' by the massive gauge
field, $\varphi$ is restored to the status of a Goldstone particle
(massless). However, if it indeed becomes the third polarization of the
massive gauge field (see section~\ref{sec:UNITARITY}), instead of being
massless, it is given, before, an infinite mass which tends to decouple it;\l
{\em Remark 2:} unlike $\xi$, $h$ cannot be reabsorbed into the massive
gauge boson.

\subsubsection{Counting the degrees of freedom.}

It is instructive, at that stage, to see by which mechanism the degrees
of freedom have been so drastically reduced, since neither the scalar field
nor the fermions are expected to be produced as asymptotic states.
We started with 2 (2 scalars) + 4 (one vector field) + 4N (N fermions)
degrees
of freedom. They have been reduced to only 3, the three polarizations of the
massive vector boson by 4N + 3 constraints which are the following:\l
- the 2 constraints linking $\varphi$ and $H$ to the fermions;\l
- the gauge fixing needed by gauge invariance;\l
- the 4N constraints coming from the condition
$\langle\ol\Psi\Psi\rangle =N\mu^3$: indeed because of the
underlying fermionic $O(N)$ invariance of the theory, this condition
is equivalent to
\be
 \langle\ol\Psi_n\Psi_n\rangle =\mu^3,
\  for\  n=1 \ldots\  N,
\label{eq:PSI1}\ee
itself meaning (see \cite{NovikovShifmanVainshteinZakharov})
\be
 \langle\ol\Psi ^\alpha_n\Psi^\alpha_n\rangle =\mu^3/4,
\  for\  n=1 \ldots\  N\  and \ \alpha\ =1\ \ldots\  4,
\label{eq:PSI2}\ee
which make 4N equations.

{\em Remark:} eq.~(\ref{eq:PSI1}) does not imply eq.~(\ref{eq:PSI2}); only
the
reverse is true, and we take, according to
ref.~\cite{NovikovShifmanVainshteinZakharov}, the latter as a definition of
the former.

One should not conclude that the `infinitely massive' fermions play no
physical role \cite{ApplequistCarrazone}. Indeed, as shown below, they make
the anomaly disappear and, as studied below, also trigger the
usual decays of the pion into two gauge fields.

\subsubsection{The disappearance of the anomaly and the conservation of the
gauge current.}

The infinite mass of the fermions, in addition to making them
unobservable as asymptotic states (see also \cite{BellonMachet}),
makes the theory anomaly-free. Indeed, the Pauli-Villars regularization of the
triangular diagram, which yields the (covariant) anomaly, writes, $M$ being the
mass of the regulator (see fig.~5)
\be
k^\mu \Big(T_{\mu\nu\rho}(m) - T_{\mu\nu\rho}(M)\Big) = m T_{\nu\rho}(m) -M
T_{\nu\rho}(M).
\label{eq:WARD}\ee
We have
\be
\lim _{M\rar\infty} M T_{\nu\rho}(M) = -{\cal A}(g,\sigma_\mu),
\ee
where ${\cal A}(g,\sigma_\mu)$ is the anomaly; so,  when $m\rar\infty$,
the Ward Identity eq.~(\ref{eq:WARD}) now shows that the anomaly gets canceled.
\figskip
\vbox{\hskip 2cm\epsffile{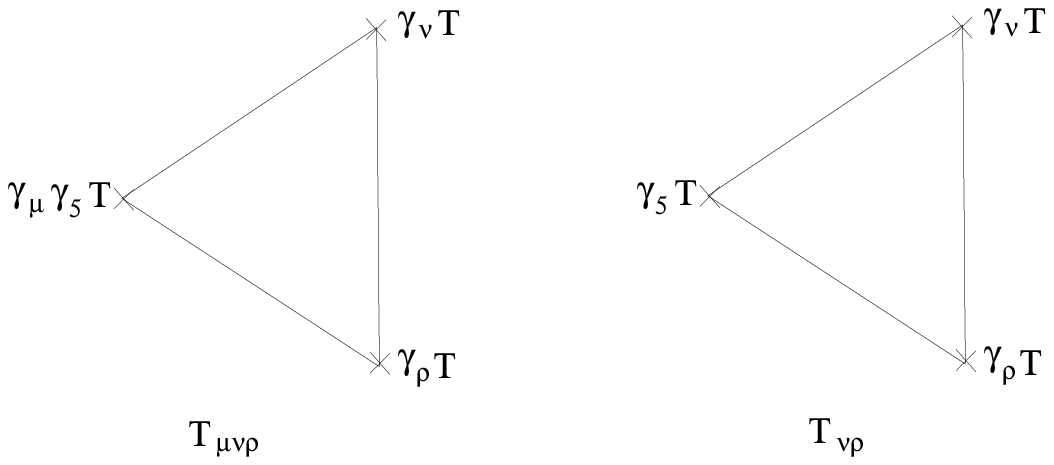}

{\centerline{\em Fig.~5: triangular diagrams involved in the anomalous Ward
Identity.}}}
\figskip
This is exactly the inverse of the situation described in
\cite{D'HokerFarhi}:
here, by decoupling, the fermions generate an effective Wess-Zumino term
exactly canceling the anomaly initially present.

The importance of the large fermion mass limit and its relevance to the
low energy or soft momentum limit has also been emphasized in
\cite{Fujikawa86} in the case of the non-linear $\sigma$-model. This case is
all the more relevant as our scalar boson has been shown
to get itself an infinite mass.

The gauge current $J_\mu^\sigma$ is conserved. It writes
\be\ba{lcl}
J_\mu^\sigma &=& g\,J_\mu^\psi + g^2\,\sigma_\mu(H^2+\varphi^2)
-g\,(\varphi\p_\mu H -H\p_\mu \varphi)\\
            &=& g\,J_\mu^\psi + g^2\, \tilde H^2 (\sigma_\mu - {1\over g}
{\p_\mu\xi\over v}).
\ea \label{eq:JSIGMA}\ee
Using the invariance of ${\cal L}_c$ by a transformation acting on both
scalars and fermions, eq.~(\ref{eq:CHVAR}), to transform the r.h.s. of the
equation below into a
variation with respect to $\xi$, the $\Psi$ equation yields
\be
\p^\mu J_\mu^\psi = v {\p{\cal L}_c\over \p\xi},
\label{eq:DJPSI}\ee
while the $\xi$ equation, deduced from the Lagrangian $\cal L$
(eq.~(\ref{eq:L1})) $+ {\cal L}_c$ (eq.~(\ref{eq:LCAB})), gives, using also
eq.~(\ref{eq:JSIGMA})
\be
\p^\mu J_\mu^\sigma -g\,\p^\mu J_\mu^\psi  = -gv\, {\p{\cal L}_c\over \p\xi},
\label{eq:DJSIGMA}\ee
Combining the two above equations we get the classical conservation of the
gauge current:
\be
\p^\mu J_\mu^\sigma = 0\label{eq:DJ}
\ee
 In the absence of anomaly, this classical equation stays
valid at the quantum level, making exact the conservation  of the gauge
current. This implements gauge invariance in the constrained theory.

\subsubsection{Retrieving the `anomalous' coupling of the pion to two gauge
fields.}\label{sec:ABANO}

The $\varphi$ into two $\sigma_\mu$'s transitions are triggered by the
coupling
of ${\cal L}_c$
\be
{i\varphi\over v} m \ol\Psi\gamma_5{\Bbb T}\Psi.
\label{eq:PHICPL}\ee
Indeed, the quantum contribution to $m \ol\Psi\gamma_5{\Bbb T}\Psi$
from the triangle precisely yields, as described in eq.~(\ref{eq:WARD}) above,
$-i\times\  the\  anomaly$, such that eq.~(\ref{eq:PHICPL}) contributes at the
one-loop level
\be
{\varphi\over v} {\cal A}(g,\sigma_\mu).
\label{eq:PHIANO}\ee
Now, after the rescaling eq.~(\ref{eq:SCALE}), eq.~(\ref{eq:PHIANO}) describes
the customary `anomalous' coupling of a neutral pion to two gauge fields
\cite{Adler}: we have
\be
{\cal A}(g,\sigma_\mu) = {\cal A}(e,a_\mu);
\label{eq:ANOANO}\ee
and, by the global rescaling by $1/a^2$ already used for the classical
Lagrangian (see sections~\ref{sec:TOYMODEL}, \ref{sec:SCALING}),
the effective coupling eq.~(\ref{eq:PHIANO}) becomes
\be
{1\over av}\pi\; {\cal A}(e,a_\mu) = {1\over f_\pi}\pi\; {\cal A}(e,a_\mu).
\label{eq:PIANO}\ee
Despite the absence of anomaly, it has been rebuilt from the constraints
and the infinite fermion mass that they yield. It appears as a consequence
of expressing $\varphi$ as a composite and of the breaking of the symmetry.

\subsubsection{Renormalizability.}

A perturbative expansion made in a theory which includes 4-fermions couplings
could be thought to be pathological. However here, the reshuffled
perturbative series built with the {\em effective} 4-fermions couplings
$\zeta(q^2)$ and $\zeta^5(q^2)$ behaves in a way that does not put
renormalizability in jeopardy. This by staying in the Nambu Jona-Lasinio
approximation.

Indeed, in this approximation, the vanishing of the effective 4-fermions
coupling constants  when $\beta \rar 0$
guarantees that all possible counterterms that could be expected in the
renormalization process at the one-loop level disappear. They are described in
fig.~6 and correspond to 6-fermions, 8-fermions, one gauge field (or one
scalar) and 4 or 6-fermions interactions; they, in addition to being
of higher order in $1/N$, go to $0$ like powers of $\beta$.
\figskip
\hskip 2cm\epsffile{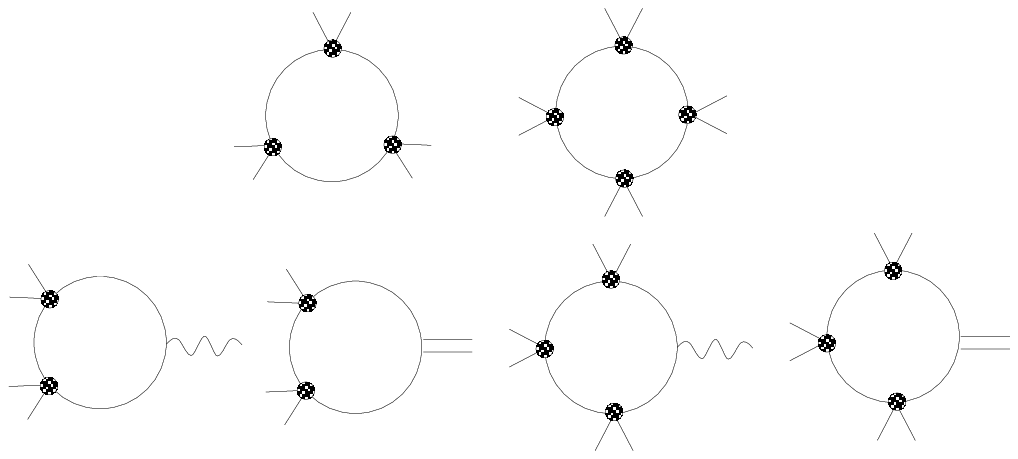}

{\centerline{\em Fig.~6: contributions to possibly uncontrollable
counterterms.}}
\figskip

Now, the natural obstacle in reshuffling a perturbative expansion is
double-counting. However, it also disappears as a consequence of the
vanishing of the diagram of fig.~7 when $\beta\rar 0$.
\figskip
\hskip 5cm\epsffile{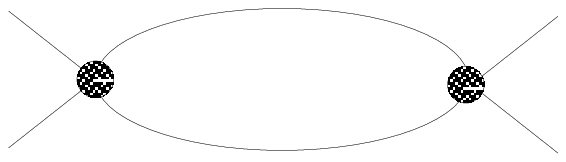}
\medskip

{\centerline{\em Fig.~7: diagram that could cause double-counting problems.}}
\figskip

While at this level, simply counting the powers of $\beta$ in the (few)
problematic diagrams is easy enough to show that the series behaves correctly,
it is clear that a demonstration of renormalizability at all orders should
rest on more powerful arguments \cite{BecchiRouetStora}, specially in the
non-abelian case; it is currently under investigation \cite{MachetThompson}.

\subsection{Back to \boldmath{$\bf SU(2)_L\times U(1)$}.}\label{sec:NAB}

Most of the previous arguments and results can be directly transposed to the
realistic non-abelian case of the Standard Model. However, some points
deserve more scrutiny:\l
- the vanishing of the anomaly and the reconstruction of the `anomalous'
couplings of the pseudoscalar mesons to the gauge fields;\l
- the fact that this theory is a theory of {\em strongly coupled gauge
fields}.

\subsubsection{Strongly coupled gauge fields.}\label{sec:STRONG}

The limit of an infinitely massive Higgs is known to go along with
a `strong' scattering of gauge bosons \cite{Veltman}. This is now consistent
with the strong interactions that pions and similar mesons undergo, as they
are precisely the third components of the massive $W$'s.
A rich resonance structure is thus expected.

The reader could worry about unitarity; it is well known that, if it
seems violated `at first order' in pion-pion scattering, it is
restored for the full amplitude; the same will occur in the physics of
$W$'s, which means in practice that unitarity, which has been proved
above at a general level, is not to be expected  order by
order in a `perturbative' expansion.

The rising of a strongly interacting sector in an originally weakly coupled
theory could provide a bridge between electroweak and strong interactions,
towards their true unification.

\subsubsection{`Anomalous' couplings between pseudoscalar mesons
and gauge fields.}\label{sec:NABANO}

I unraveled in the abelian example the mechanism which, in an anomaly-free
theory, reconstructs the `anomalous' couplings of a pion to two
gauge fields; its compositeness is the essential ingredient,
together with the infinite mass of the quarks, which also yields the
absence of any anomaly for the gauge current. The `decoupling'
\cite{ApplequistCarrazone} of the
infinitely massive fermions is thus not total, since, at the quantum (1
loop) level, the `anomalous' coupling springs out.

We learn from this example that we can only expect these kind of `quantum'
couplings for those of the scalar fields that have explicitly been
considered to be composite, {\em i.e.} for which constraints have been
introduced. In the GSW model, this means that only the triplet $\vec\varphi$
will have anomalous couplings. This is the first of our results.
 Detecting similar couplings for the
neutral $K$ or $D$ mesons would mean that our model is
still incomplete and needs an extension, in particular to include more
'constraints', which means that more mesons have to be explicitly taken
composite, meaning itself probably that the massive gauge field structure is
richer than that of a sole  triplet of $W$'s.

Next, the specificity of the three pseudoscalar fields $\vec\varphi$,
which in particular involve the Cabibbo angle $\theta_c$, gives to  these
couplings precise expressions which also involve the mixing angle. Phrased in
another way, the embedding of the gauge group into the chiral group controls
the form of the anomalous couplings of the charged pseudoscalar mesons.
However, as those automatically involve one charged, and thus massive, gauge
field,  the experimental verification of this dependence on the mixing angle
is not trivial.

The Lagrangian of constraint ${\cal L}_c$ writes now
\be
{\cal L}_c= \lim_{\beta\rar 0}\frac{-N\Lambda^2}{2\beta}
\left(H^2 -{\vec\phi}^{\:2} -{2v\over N\mu^3}(H{\ol\Psi\Psi\over\sqrt{2}}-
\sqrt{2}\vec\phi.\ \ol\Psi\gamma_5 \vec{\Bbb T} \Psi)+
 {v^2\over N^2\mu^6}\Bigl({(\ol\Psi\Psi)^2\over 2}
-2(\ol\Psi\gamma_5 \vec{\Bbb T} \Psi)^2\Bigr)\right),
\label{eq:LC}
\ee
and the rescaling of the fields goes, in analogy with eq.~({\ref{eq:SCALE}),
 according to
\be
P^i = {v\over 2 f_\Pi} \varphi^i = c^i_\alpha\,\Pi^\alpha,
\label{eq:RESCALE}\ee
with the following conventions:\l
$i$ spans the set of three indices $(+,-,3)$ corresponding to the three
$SU(2)$ generators ${\Bbb T}^+, {\Bbb T}^-, {\Bbb T}^3$, and $\alpha$ spans
the set of $N^2$ indices of $U(N)$;
\be
{\Bbb T}^i = c^i_\alpha {\Bbb T}^\alpha;
\ee
${\Bbb T}^\alpha$ is a $U(N)$ generator, corresponding to the `strong'
pseudoscalar eigenstate $\Pi^\alpha$ (considered here as a generic name for
pions, kaons \ldots). The coefficients $c^i_\alpha$ depend of the mixing
angle. $f_\Pi$ is the coupling constant of the pseudoscalars, taken, for the
sake of simplicity, to be the same for all of them.

The rescaled Lagrangian ${{v^2\over 4{f_\Pi}^2}\cal L}_c$
involves the couplings
\be
{v\over\sqrt{2}{f_\Pi}^2} m_0\sum_{i}\varphi^i \ol\Psi\gamma_5 {\Bbb T}^i \Psi
=
{\sqrt{2}\over f_\Pi} m_0 \sum_{i,\alpha,\beta} c^i_\alpha c^i_\beta
\;\Pi^\alpha\; \ol\Psi \gamma_5 {\Bbb T}^\beta \Psi,
\label{eq:NABPICPL}\ee
where the quark mass $m_0$ is
\be
m_0 = -\frac{\Lambda^2 v^2}{2\beta \mu^3}.
\ee
The `anomalous' coupling of $\Pi^\alpha$ to the generic gauge fields
$A_\mu^a$ and $A_\mu^b$, themselves coupled to the quarks by the
matrices ${\Bbb T}^a \gamma^\mu$ and ${\Bbb T}^b \gamma^\mu$ or ${\Bbb T}^a
\gamma^\mu\gamma_5$ and ${\Bbb T}^b  \gamma^\mu\gamma_5$ occur via the
triangle diagram depicted in fig.~8:
\figskip
\hskip 2cm \epsffile{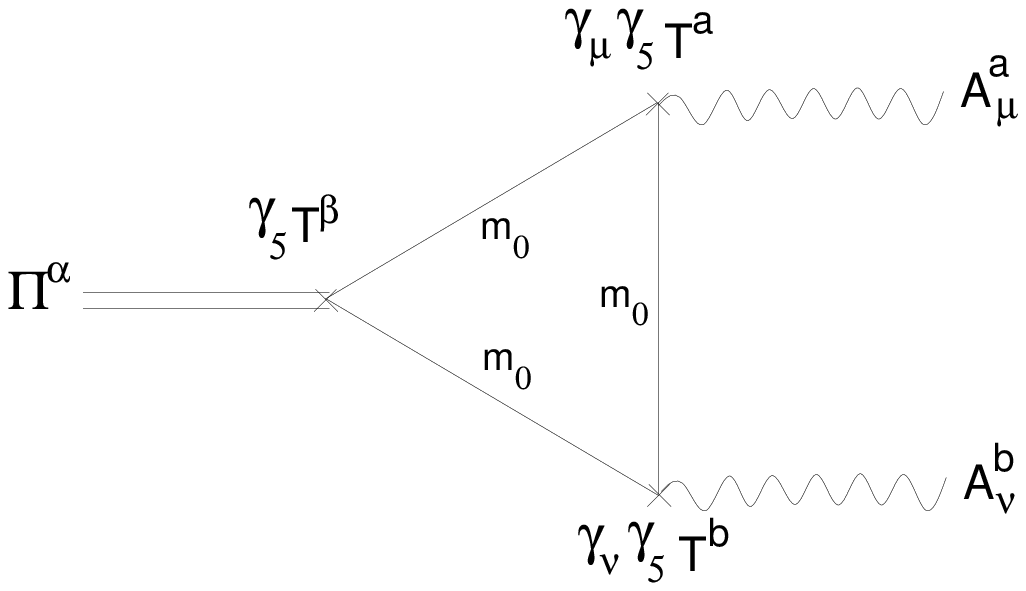}

{\centerline{\em Fig.~8: `anomalous' coupling of a pseudoscalar to two gauge
fields.}}
\figskip
For infinitely massive fermions, it yields the covariant form of the
anomaly, and one gets the coupling
\be
\eta\;{\sqrt{2}\over f_\Pi} \Pi^\alpha\;\sum_{i,\beta} c^i_\alpha c^i_\beta \;
Tr\; ({\Bbb T}^\beta \{{\Bbb T}^a ,{\Bbb T}^b\})\;F_{\mu\nu}^a F^{\mu\nu b},
\label{eq:ANOCPL}\ee
where $\eta$ is the normalization factor, and the $F_{\mu\nu}$'s are the
field strengths associated with the $A_\mu$ gauge fields.

We recall that only triangles with one or three $\gamma_5$ matrices will
yield such couplings; as usual, the group $SU(2)$ being such that, for any
set of three matrices $t_1,t_2,t_3$ belonging to it, we have
$Tr\; t_1\{t_2,t_3\} = 0$, only triangles involving `mixed' couplings will
yield non-vanishing results (a typical case being the $\pi^0$ to
$\gamma\gamma$ coupling).

{\em Remark 1:} the normalization factor of the {\em covariant} form of the
anomaly is known to be three times that of the {\em consistent} anomaly
\cite{BardeenZumino}.
Ours is thus not automatically the same as that of the original work of Wess
and Zumino;\l
{\em Remark 2:} the same remark as in section~\ref{sec:ABANO} concerning the
loop expansion with rescaled fields is of course valid here.

To conclude this section, I outline  a trivial consequence of
eq.~(\ref{eq:ANOCPL}): the couplings of the $\pi^+$ and $K^+$ mesons to
two gauge fields are in the ratio $\cos\theta_c/\sin\theta_c$. Other similar
relations can of course be immediately deduced.

\section{Conclusion.}

This work has emphasized conceptual differences which could increase
at the lowest cost the predictive power of the Standard Model. I hope to have
convinced the reader that adding {\em very little} information may have
quite large consequences.

I, maybe, did not insist enough on the economy of this approach:
both here and in \cite{BellonMachet}, we are able to
describe many aspects of electroweak physics ({\em e.g.} the absence of a
right-handed neutrino, the $V-A$ structure of the leptonic weak currents, the
non-observation of the quarks, the existence of several mass scales,
degeneracies in the mass spectrum of scalar and pseudoscalar mesons,
 their `anomalous' decays into two photons \ldots),
without destroying any basic concept (the Standard Model stays practically
untouched),  nor predicting a jungle of new particles.
It is true that I did not investigate all couplings of the mesonic sector,
and that further studies may yield surprises and
possible precise experimental tests.  This is of course under scrutiny (see
in particular ref.~\cite{Machet1}).
Also, the first prediction stays a negative one, which states that the Higgs
boson will not be observed.

The absence of  baryons only reflects my inability to incorporate them
into a coherent dynamical framework.
I also deliberately postponed the study of vector mesons.

One can argue that I did not investigate the corrections
that appear beyond the Nambu Jona-Lasinio approximation, and that they
could put renormalizability in jeopardy. This is another important
aspect which is currently under investigation. It must however be stressed
that impressive amounts of physics can be and have been done with
effective and
non-renormalizable Lagrangians ({\em e.g.} the Fermi theory of weak
interactions, chiral Lagrangians, $\sigma$-models \ldots) which have there
own interest and relevance, and are important steps towards  more fundamental
theories.

As I scattered in the core of the paper numerous remarks concerning other
models,
further investigations, various hopes and goals, I will not lengthen this
conclusion, but only suggest again that the emergence of a strongly interacting
sector in one-to-one correspondence with particles which, we know, do undergo
strong interactions, could be a hint for the direction to take towards a
true unification of strong and electroweak interactions.

\bigskip {\em\underline {Acknowledgments}: it is a pleasure to thank
J. Avan, O. Babelon, M. Bellon, G. Rollet, M. Talon, \hbox{G. Thompson}  and
\hbox{C.-M. Viallet} for their advice, help, and encouragement.}

\newpage\null
\listoffigures
\bigskip
\begin{em}
Fig.~1:  diagram generating the leptonic decays of $\varphi$;\l
Fig.~2:  the effective 4-fermions coupling $\zeta(q^2)$;\l
Fig.~3:  resumming the fermion propagator;\l
Fig.~4:  resumming the scalar propagator;\l
Fig.~5:  triangular diagrams involved in the anomalous Ward Identity;\l
Fig.~6:  contributions to possibly uncontrollable counterterms;\l
Fig.~7:  diagram that could cause double-counting problems;\l
Fig.~8:  `anomalous' coupling of a pseudoscalar to two gauge fields.

\newpage\null

\end{em}
\end{document}